\documentclass[12pt]{article}

\usepackage{amscd}

\usepackage{verbatim}

\usepackage{amssymb}

\usepackage{amsmath}

\newcommand{\be}{\begin{equation}}
\newcommand{\ee}{\end{equation}}
\newcommand{\ben}{\begin{eqnarray}}
\newcommand{\een}{\end{eqnarray}}
\newcommand{\ba}{\begin{eqnarray}}
\newcommand{\ea}{\end{eqnarray}}
\newcommand{\nn}{\nonumber \\}
\newcommand{\bi}{\begin{itemize}}
\newcommand{\ei}{\end{itemize}}

\newcommand{\lb}{\left (}
\newcommand{\rb}{\right )}
\newcommand{\ltb}{\left [}
\newcommand{\rtb}{\right ]}

\newcommand{\p}{\partial}
\newcommand{\ra}{\rightarrow}

\renewcommand{\theequation}{\thesection.\arabic{equation}}

\begin{document}
\newcommand{\diff}{\frac{d}{dr}}
\newcommand{\ddiff}{\frac{d^2}{dr^2}}
\newcommand{\Rdiff}{\frac{d}{dR}}
\newcommand{\Rddiff}{\frac{d^2}{dR^2}}
\begin{center}

\vspace{24pt} { \large \bf Non-spherically symmetric black string perturbations in the large D limit} \\

\vspace{30pt}

\vspace{30pt}

\vspace{30pt}

{\bf Amruta Sadhu\footnote{sadhuamruta@students.iiserpune.ac.in}}, {\bf Vardarajan
Suneeta\footnote{suneeta@iiserpune.ac.in}}

\vspace{24pt} 
{\em  The Indian Institute of Science Education and Research (IISER),\\
Pune, India - 411008.}

\end{center}
\date{\today}
\bigskip

\begin{center}
{\bf Abstract}
\end{center}
We consider non-spherically symmetric perturbations of the uncharged black string/flat black brane in the large dimension (D) limit of general relativity. We express the perturbations in a simplified form using variables introduced by
Ishibashi and Kodama. We apply the large D limit to the equations, and show that this leads to decoupling of the equations in the near-horizon and asymptotic regions. It also enables use of matched asymptotic expansions to obtain approximate analytical solutions and to analyze stability of the black string/brane. For a large class of non-spherically symmetric perturbations, we prove that there are no instabilities in the large D limit. For the rest, we provide additional matching arguments that indicate that the black string/brane is stable. In the \emph{static} limit, we show that for \emph{all} non-spherically symmetric perturbations, there is no instability. This is proof that the Gross-Perry-Yaffe mode
for semiclassical black hole perturbations is the unique unstable mode even in the large D limit. This work is also a direct analytical indication that the only instability of the black string is the Gregory-Laflamme instability.

\newpage
\section{Introduction}
Perturbative stability of black strings and black branes is an important issue in gravitation and string theory. Since the discovery of the instability of the black string in dimension greater than four by Gregory and Laflamme (GL) \cite{GL}, \cite{GL1}, research has focussed on understanding questions such as the endpoint of the GL instability (an account
of this can be found in \cite{kolreview}). The GL instability for the five dimensional black string is related to the semiclassical instability found by Gross, Perry and Yaffe (GPY) \cite{gpy} of the four dimensional Schwarzschild instanton after a suitable gauge choice \cite{reall}. Similar instabilities have been found in the analysis of perturbations of black branes (see \cite{obers} and references therein). Much of this work has centered on \emph{spherically symmetric} perturbations of black strings and black branes, as the equations are more tractable, and the GL instability falls in this category (for a review, see \cite{berti}). The perturbation equations for non-spherically symmetric perturbations are coupled and it is not possible to solve them analytically in general. Due to a study of the link
between local thermodynamic instability and classical instability of extended objects \cite{GM}, \cite{GM1} \cite{reall},
\cite{hirayama}, \cite{hovdebo}, \cite{miyamoto}, it is expected that non-spherically symmetric perturbations do not cause instabilities (see section 6, \cite{obers} for a review of the correlated stability conjecture)\footnote{An analogy of the GL instability to the Rayleigh-Plateau instability of fluids has also been used to argue this.\cite{cardoso}}. However, there exists no fully analytical proof of this even for the asymptotically flat black string in general dimensions, due to the difficulty in analyzing the coupled non-spherically symmetric perturbations. Furthermore, the study of the evolution of \emph{stable} non-spherically symmetric perturbations, particularly quasinormal modes, is also of interest in physics.

In this paper, we obtain the first breakthrough in the analysis of the coupled equations corresponding
to non-spherically symmetric perturbations of black strings and branes. This is achieved in the large dimension ($D$) limit of general relativity which was first employed in \cite{kol} to study the spherically symmetric GPY mode for the Schwarzschild instanton. A bigger framework for the large dimension limit, with applications to perturbations of black holes/branes has been developed by Emparan, Suzuki and Tanabe \cite{emparan}. We apply this limit, for the first time, to the \emph{non-spherically symmetric} perturbations of the black string/flat black brane. Using this, we provide the first analytical proof of the stability of the black string/flat black brane under a vast class of non-spherically symmetric perturbations. Classifying perturbations based on their decomposition in terms of scalar, vector and tensor spherical harmonics, the evolution of the tensor mode is the easiest to analyze as it can be reduced to a single ordinary differential equation (ODE). It has been shown by Kodama to not lead to instabilities \cite{kodamanotes}.\footnote{Kudoh \cite{kudoh} has partly analyzed non-spherically symmetric perturbations of the black string. However the analysis has incorrect equations and claims that do not agree both with our
results and related work of Kodama \cite{kodamanotes} (see also \cite{vstensor}) and Gibbons and Hartnoll \cite{gibhart}. This is discussed in Appendix A.} It is the vector and scalar modes which are difficult to analyze analytically, since each set of perturbations involves many coupled equations. To analyze these modes, we choose a gauge such that perturbations with an index on the brane/extra dimension vanish (this was used by Reall \cite{reall} to prove equivalence of the GL and GPY modes). The perturbations are then expressed in terms of variables introduced by Ishibashi and Kodama \cite{rev}, \cite{iks}. The vector perturbations reduce to a set of two coupled ODEs, and the non-spherically symmetric scalar perturbations are a set of three coupled ODEs. Both sets of equations do not decouple even in the large $D$ limit. However, they decouple in the near-horizon and asymptotic regions. By using matched asymptotic expansions, we prove that the vector equations and part of the scalar equations do not lead to instabilities. We perform an additional matching procedure to argue that the rest of the scalar perturbations are also not likely to lead to instabilities. In the static limit, we show that for \emph{all} classes of non-spherically symmetric perturbations, there is no instability. This is a proof that the GPY mode is the unique unstable mode of semiclassical perturbations of the Schwarzschild-Tangherlini black hole
in the large $D$ limit. It is also direct evidence that the Gregory-Laflamme instability is the unique instability for the uncharged black string. Furthermore, these techniques provide approximate analytical solutions to the perturbation equations and can also be used to obtain quasinormal modes corresponding to non-spherically symmetric perturbations.

In the case of spherically symmetric perturbations, quasinormal modes of black strings have been analyzed in \cite{konoplya}. In the context of anti-de Sitter holography, quasinormal modes corresponding to scalar field perturbations and non-spherically symmetric perturbations of branes in anti-de Sitter spacetime have been discussed in \cite{starinets1}, \cite{starinets2}. The large $D$ limit has been used to discuss quasinormal modes and instabilities of black holes in \cite{qnmemparan,rotatingemparan,qnm2emparan, qnmadsemparan}. Other applications of the large $D$ limit to the dynamics of
black holes, black rings and black branes include \cite{grumiller,konoplya1,santos,suzuki,suzuki2,tanabe1,tanabe2,chen}. An effective description of black holes in the large $D$ limit by surfaces has been explored in \cite{effectiveemparan}. An interesting direction is the study of nonlinear evolution of perturbations of black strings/branes in the large $D$ limit \cite{nlemparan, nlizumi, nlminwalla, nlminwalla1}.

The plan of the paper is as follows: in section II, we outline the methodology. We discuss the gauge-fixing procedure employed, the perturbation equations, certain perturbation variables defined by Ishibashi and Kodama, as well as their application to our problem. In section III, we discuss vector perturbations. We derive the vector perturbation equations in a simplified form and analyze them in a large $n$ limit, where $n$ is the dimension of the $n$-sphere part of the metric. We discuss matching of solutions from the near-horizon and the
asymptotic regions. In section IV, we do this procedure to the non-spherically symmetric scalar perturbations. Section V
is a summary of our results, and section VI is a brief discussion of future projects. There are four appendices with lengthy calculations used in the paper: Appendix A discusses tensor perturbations. Appendix B discusses how source terms in some of the equations are handled. Appendix C derives in great detail, a very crucial part of our paper: the simplified non-spherically symmetric scalar equations. Appendix D discusses a certain asymptotic expansion of modified Bessel functions which is used in many places.

\section{Non-spherically symmetric perturbations of the black string: Methodology}
In this section, we will outline the strategy for the analysis of the non-spherically symmetric perturbations of the black string/flat black brane. We will also summarize the tools required. For simplicity of notation, we will discuss the black string --- the same computation extends to the flat black brane as well.

The (uncharged) black string metric is $D = n+3$ dimensions, obtained by adding a flat extra dimension to an $n + 2$ dimensional Schwarzschild-Tangherlini metric is
\begin{equation}
g_{\mu\nu}dx^\mu dx^\nu= - f(r) dt^2 + f^{-1} (r)dr^2 +r^2 d\Omega^{2}_{n} + dz^2;
\label{giv1}
\end{equation}
where $f (r) = \left(1-\frac{b^{n-1}}{r^{n-1}}\right)$ and $r > b$ ($r=b$ is the location of the horizon). In the case of a flat black brane of dimension $D=n + 2 + p$, the flat metric corresponding to $p$ extra dimensions is added. \\
We will use capital Roman indices $A,B,...$ to denote coordinates on the black string (or brane). Greek indices $\mu, \nu,...$ will be used to denote indices only in the Schwarzschild-Tangherlini part of the metric, and coordinates in this part of the metric will be denoted collectively by $y$.
We consider perturbations of the metric (\ref{giv1}), with the perturbed metric $\bar{g}_{AB}=g_{AB}+ \bar h_{AB}$, in linearized perturbation theory. We first make a gauge choice used by Reall \cite{reall} that allows us to set $\bar h_{A z} = 0$. Similarly, for the flat black brane, this can be used to set all metric perturbations with an index on the brane to zero. Reall used this gauge to study scalar s-wave perturbations, but it can be used for all black brane perturbations, and the only non-zero perturbations left after gauge-fixing are thus the ones with indices in the Schwarzschild part of the metric. Furthermore, as showed in \cite{reall}, some of the linearized Einstein equations then imply that these perturbations have to be transverse and traceless. The linearized Einstein equation for the black string perturbations is
\begin{equation}\label{EE:n+3}
\delta R_{MN}=0
\end{equation}
The linearized Ricci tensor $\delta R_{MN}$ is expressed in terms of the Lichnerowicz Laplacian $\Delta_L$ acting on the perturbations as
\begin{eqnarray}\label{lin-ricci}
&2\delta R_{MN}= \Delta_L \bar h_{MN}-\nabla_M\nabla_N \bar h+\nabla_M\nabla_S \bar h_N^S+\nabla_N\nabla_S \bar h_{M}^S  \\[0.2 cm]
&\Delta_L \bar h_{MN}=-\nabla^L\nabla_L \bar h_{MN}+R_{ML}\bar h^L_N+R_{NL}\bar h^L_M-2R_{MLNS}\bar h^{LS}.
\end{eqnarray}
All curvature tensors are those of the black string metric (\ref{giv1}). $\bar h = g^{MN}\bar h_{MN}$.
For this background metric, the Laplacian acting on symmetric tensors splits in the form
\begin{equation}
\nabla^L\nabla_L=\nabla^\mu\nabla_\mu + \partial_z^2.
\end{equation}

For the metric (\ref{giv1}), the gauge choice $\bar h_{Mz} = 0$ reduces (\ref{EE:n+3}) to
\begin{equation}\label{EE:n+2}
\Delta_L^{Sch} \bar h_{\mu\nu}-\nabla_\mu\nabla_\nu \bar h+\nabla_\mu\nabla_\sigma \bar h_\nu^\sigma+\nabla_\nu\nabla_\sigma \bar h_\mu^\sigma = \partial_z^2 \bar h_{\mu\nu}
\end{equation}

$\Delta_L^{Sch} \bar h_{\mu\nu}$ denotes the Lichnerowicz Laplacian of the Schwarzschild-Tangherlini metric acting on perturbations of this metric. Following Gregory and Laflamme, we choose  the ansatz for $\bar h_{\mu\nu} (y, z)$
\begin{equation}\label{fourier-z}
\bar h_{\mu\nu} (y,z) =e^{i\lambda z}h_{\mu\nu}(y).
\end{equation}
(\ref{EE:n+2}) then becomes
\begin{equation}\label{giv1a}
\Delta_L^{Sch} h_{\mu\nu}-\nabla_\mu\nabla_\nu h+\nabla_\mu\nabla_\sigma h_\nu^\sigma+\nabla_\nu\nabla_\sigma h_\mu^\sigma = -\lambda^2h_{\mu\nu}
\end{equation}
Here, $h = g^{\mu \nu} h_{\mu \nu}$.
As emphasized by Reall \cite{reall}, we have already fixed gauge, but now the other Einstein equations with indices on the brane can be used to show that $h_{\mu \nu}$ must be transverse and traceless. Putting this back in (\ref{giv1a}), we get
\begin{equation}\label{giv1bricci}
\Delta_L^{Sch} h_{\mu\nu} = -\lambda^2 h_{\mu\nu}.
\end{equation}
Thus, we finally obtain an eigenvalue equation for the Lichnerowicz Laplacian in the Schwarzschild-Tangherlini background. Negative eigenvalues corresponding to normalizable eigentensors (normalizable with respect to the volume form of the background) are relevant for perturbations of the black string. For the black brane, the generalization of the ansatz (\ref{fourier-z}) is $\bar h_{\mu\nu}=e^{i\lambda_{k} z^{k}}h_{\mu\nu}(y)$ where $k$ runs from $1$ to $p$ (the number of extra dimensions) and $\lambda^2 = \Sigma_{k=1}^{p} \lambda_{k}^{2} $. Setting $\lambda = 0$ gives the equation for classical perturbations of the Schwarzschild-Tangherlini black holes, and Ishibashi and Kodama have already proved their stability \cite{rev}. To explore stability of black strings/branes, the nontrivial case to analyze is solutions to (\ref{giv1bricci}) with $\lambda \neq 0$. For future reference, we note that for a transverse, traceless $h_{\mu \nu}$, (\ref{giv1bricci}) is equivalent to
\begin{equation}
\delta G_{\mu \nu} = - \frac{1}{2} \lambda^2 h_{\mu \nu};
\label{giv1b}
\end{equation}
where $\delta G_{\mu \nu} $ is the first variation of the Einstein tensor evaluated for the transverse traceless perturbation $h_{\mu \nu} $ on the Schwarzschild-Tangherlini background. Equivalently, we could have set the linearized Einstein tensor for the brane to zero and obtained (\ref{giv1b}).

(\ref{giv1b}) is a set of many coupled equations for the perturbations $h_{\mu \nu}$. The perturbations can be decomposed in terms of scalar, vector and tensor (for $n > 2$) spherical harmonics  on the $n$-sphere with metric $d\Omega^{2}_{n}$ --- each class (which we call scalar, vector and tensor perturbations, respectively) decouples and can be studied separately. Tensor perturbations have already been discussed in \cite{kodamanotes} --- they do not lead to instability. The vector and scalar perturbations are each solutions of intricately coupled equations.

To analyze the vector and scalar perturbation equations, we will adapt a formalism due to Ishibashi and Kodama (IK) originally developed for studying classical gravitational perturbations of Schwarzschild-Tangherlini spacetimes \cite{rev}. Since the linearized Ricci tensor is invariant under a gauge transformation, for $\lambda = 0$, (\ref{giv1b}) is invariant. Ishibashi and Kodama introduced manifestly gauge-invariant variables by taking suitable combinations of metric perturbations of the Schwarzschild spacetime. The linearized Ricci tensor can be written entirely in terms of these variables. We will use their variables even for $\lambda \neq 0$, and take appropriate combinations of the various equations to obtain equations for the black string perturbations written entirely in terms of the IK variables. This is done mainly for computational simplicity. However, both in the vector and scalar case, we are still left with coupled ODEs, and in subsequent sections, we will employ the large $D$ limit, as well as the method of matched asymptotic expansions to analyze them.

\vskip 0.3cm
{\bf The Ishibashi-Kodama variables}
\vskip 0.3cm
 We will set up the notation for our paper by quickly stating the perturbation variables proposed by Ishibashi and Kodama in (\cite{rev}, \cite{iks}) for doing gravitational perturbation theory. Their variables are useful for studying perturbations of the Schwarzschild-Tangherlini metric
\begin{eqnarray}
g_{\mu\nu}dy^\mu dy^\nu &=& g_{ab}(x)dx^a dx^b+r^2(x) d\Omega^2_n ; \nonumber \\
&=& - f(r) dt^2 + f^{-1} (r) dr^2 +r^2 d\Omega^{2}_{n}.
 \label{giv2}
\end{eqnarray}

Since, after gauge-fixing, we also work with perturbations of the metric (\ref{giv2}), we will use the IK variables.
Here $g_{ab}(x)$ is the $r-t$ part of the metric and $d\Omega^2_n=\gamma_{ij} d\tilde{y}^i d\tilde{y}^j$ is the metric of a $n$-dimensional sphere of unit radius with Ricci tensor given by $\hat{R}_{ij}=(n-1) \gamma_{ij}$.

We use indices $a,b$ to denote indices from the set $r,t$ and indices $i,j$ of coordinates on sphere. Indices $\mu,\nu$ denote any coordinate in the spacetime with metric (\ref{giv2}). Covariant derivatives and Ricci tensors on each space are denoted as
\begin{align*}
&g_{\mu\nu} \rightarrow \nabla_\mu , R_{\mu\nu}\\
&g_{ab} \rightarrow D_a , {^m}R_{ab}\\
&\gamma_{ij} \rightarrow \hat{D}_i, \hat{R}_{ij}.
\label{giv3}
\end{align*}
We consider perturbations of the metric (\ref{giv2}) with the perturbed metric denoted by $g_{\mu\nu}^{p}=g_{\mu\nu}+h_{\mu\nu}$ in linearized perturbation theory, where $h_{\mu \nu}$ is defined in terms of the original black string perturbation by (\ref{fourier-z}). The scalar, vector and tensor components of $h_{\mu \nu}$ are defined as those that are decomposed in terms of scalar, vector and tensor spherical harmonics on the $n$-sphere, respectively. The components $h_{ab}$ are scalars with respect to transformations on the $n$-sphere. The other components can be written as follows:
\begin{align}
 &h_{ai}=\hat{D}_i h_a+ h^{(1)}_{ai}\\
 &h_{ij}={h_{T}^{(2)}}_{ij}+2\hat{D}_{(i)}{h^{(1)}_T}_{j}+h_L\gamma_{ij}+\hat{L}_{ij}h_T^{(0)}.
\label{giv4}
\end{align}
 where
 \begin{align}
&\hat{D}^j{h_{T}^{(2)}}_{ij}={h_{T}^{(2)i}}_i=0\\
&\hat{D}^a h^{(1)}_{ai}=0, \hat{D}^j{h_{T}^{(1)}}_j=0.
\label{giv5}
 \end{align}
Here ${h_{T}^{(2)}}_{ij}$ is the `tensor' part, the `vector' set is (${h_{T}^{(1)}}_j,h^{(1)}_{ai}$), and the `scalar' set is ($h_{ab},h_a,h_L,h_T^{(0)}$). The eigenvalue equations for the linearized Ricci tensor decouple for these three classes, and they can be studied separately.

Ishibashi and Kodama consider combinations of perturbations in each set which are gauge invariant. To do this, the generator of a gauge transformation $\xi_\mu$ is also decomposed into a `vector' part and a gradient of a scalar.
We recall that under a gauge transformation generated by any infinitesimal vector $\xi_\mu$, the perturbation transforms as
\begin{equation}
h_{\mu\nu}^{'} = h_{{\small \mu\nu}}-\nabla_\mu\xi_\nu-\nabla_\nu\xi_\mu .
\label{giv6}
\end{equation} \\

\textbf{Vector perturbations}:  We use the notation of \cite{rev}, \cite{iks} to describe the vector perturbations.\footnote{ Ishibashi and Kodama use the notation $f_a$ and $H_T$ to denote distinct quantities in the vector and scalar case. Here we add a superscript $vector$ in the vector case to avoid confusion.}
\begin{align}
h_{ab}=0 \qquad h_{ai}=r f_{a}^{vector} V_i \qquad h_{ij}=2r^2H_{T}^{vector}V_{ij}.
\label{giv7}
\end{align}
Here $f_{a}^{vector},H_{T}^{vector}$ are functions of $r, t$. Vector harmonics $V_i$ and $V_{ij}$ are defined by
\begin{align}
(\hat{\Delta}+k_{v}^2)V_i=0, \hat{D}_iV^i=0\\
V_{ij}=-\frac{1}{2k_{v} }(\hat{D}_iV_j+\hat{D}_jV_i).
\label{giv8}
\end{align}
$k_{v}^2=l(l+n-1)-1$ and $l=2,... $.  We denote $\hat{\Delta} = \gamma^{ij}\hat{D}_i \hat{D}_j $. Gauge-invariant variables in the class of vector perturbations are given by the combination
\begin{equation}\label{Fa}
F_{a}=f_{a}^{vector}+\frac{r}{k_v}D_a H_{T}^{vector}
\end{equation}
\\
\textbf{Scalar perturbations}:
Similarly one can construct gauge-invariant variables for scalar perturbations. Scalar perturbations are given by \cite{iks}, \cite{rev}
\begin{align}
h_{ab}=f_{ab}S \qquad h_{ai}=rf_aS_i \qquad h_{ij}=2r^2(H_L\gamma_{ij}S+H_TS_{ij})
\end{align}
$S, S_i$ and $S_{ij}$ are scalar harmonics satisfying
\begin{align*}
&(\hat{\Delta}+k^2)S=0 \qquad S_i=-\frac{1}{k}\hat{D}_iS \qquad \hat{D}_iS^i=kS \\
& S_{ij}=\frac{1}{k^2}\hat{D}_i\hat{D}_jS+\frac{1}{n}\gamma_{ij}S \qquad S^i_i=0
\end{align*}
$k^2=l(l+n-1)$ and $l=0,1,2...$. We will not consider $l=0$ which corresponds to spherically symmetric perturbations, since this case has already been extensively analyzed by Gregory and Laflamme and the GL instabilities fall in this class. The modes with $k^2 = n$ (i.e., $l=1$) are exceptional modes, in the sense that the construction of gauge-invariant variables is not possible in this case (for details, see \cite{iks}). We will eventually work in a large $n$ approximation where it is not possible to consider the exceptional mode. Therefore, for the discussion that follows on scalar modes, we will consider only $ l \geq 2$.
Gauge invariant variables for scalar perturbations (not defined for $l=0$ and $l=1$)are constructed as follows:
First we define
\begin{equation}
X_a=\frac{r}{k}\biggl(f_a+\frac{r}{k}D_aH_T\biggr)
\end{equation}
In terms of $X_a$, the gauge invariant variables are
\begin{align}\label{Fab}
& F_{ab}=f_{ab}+D_aX_b+D_bX_a\\
& F=H_L+\frac{1}{n}H_T+\frac{1}{r}D^aX_a
\end{align}

In the next two sections, we will consider the equations for the vector and scalar perturbations arising from (\ref{giv1b}), written using IK variables. To analyze the coupled equations in each case, we will use the large $n$ approximation.

\section{Vector Perturbations}
In this section, we will look at solutions to (\ref{giv1b}) for the class of vector perturbations (\ref{giv7}). Our goals are two-fold: (i) to prove that the black string (brane) is stable under this class of perturbations. (ii) to develop approximate solutions for vector perturbations.

We can write the equations (\ref{giv1b}) in terms of the IK variables $F_a$ defined in the previous section (\ref{Fa}). For the expression of the variation of the Einstein tensor in terms of these variables, we refer the reader to \cite{iks}, \cite{rev}.
Upon using these results, the equations $\delta G_{ai}= - \frac{1}{2} \lambda^2 h_{ai}$ and $\delta G_{ij}= - \frac{1}{2} \lambda^2 h_{ij}$ are written in terms of the IK variables as
\begin{align}
&\frac{1}{r^{n+1}}D^b\left[r^{n+2}\left[D_b\left(\frac{F_a}{r}\right)-D_a\left(\frac{F_b}{r}\right)\right]\right]- \frac{\alpha}{r^2}F_a=\lambda^2 f_{a}^{vector} \nonumber \\
&\frac{k_v}{r^n}D_a(r^{n-1}F^a)=\lambda^2 H_{T}^{vector};
\label{vector1}
\end{align}
where $\alpha = k_{v}^{2}-(n-1)$.
\\We can use the equations (\ref{vector1}) to obtain second order differential equations for the variables $F_a$ (i.e., $F_r$ and $F_t$).
\begin{align}\label{main}
&\square F_a - D^bD_aF_b+ D_aD^bF_b + n\frac{D^brD_bF_a}{r} - 2\frac{D^brD_aF_b}{r} -\frac{\square r}{r}F_a  - n\frac{(Dr)^2}{r^2}F_a \nonumber \\
& - (n-2)\frac{D^brD_ar}{r^2}F_b + \frac{D^bD_ar}{r}F_b + (n-1)\frac{D_aD^br}{r}F_b - \frac{\alpha}{r^2}F_a=\lambda^2 F_a
\end{align}

Explicitly evaluating the covariant derivatives, we get a system of coupled equations for $F_r$ and $F_t$.
\begin{equation}
f\partial_r^2F_t-\frac{1}{f}\partial_t^2F_t+\frac{nf}{r}\partial_rF_t-\left[\frac{nf}{r^2}+
\frac{\alpha}{r^2}\right]F_t+\left[f'-\frac{2f}{r}\right]\partial_tF_r=\lambda^2 F_t
\label{Ft}
\end{equation}
\begin{align}
&f\partial_r^2F_r-\frac{1}{f}\partial_t^2F_r+\left[2f'+\frac{(n-2)f}{r}\right]\partial_rF_r\nonumber \\
&\qquad\quad+\left[f''+\frac{(n-2)f'}{r}-\frac{2(n-1)f}{r^2}-\frac{\alpha}{r^2}\right]F_r + \frac{f'}{f^2}\partial_tF_t=\lambda^2 F_r
\label{Fr}
\end{align}

The equations (\ref{Ft}) and (\ref{Fr}) we have obtained match with those of Kodama\cite{kodamanotes}. Kodama has an additional equation owing to the fact that \cite{kodamanotes} employs gauge-invariant variables in the entire black brane, while we work with gauge-fixed variables in the black hole spacetime.
We first do a modal decomposition of $F_t$ and $F_r$.
\begin{align}\label{modalF}
F_t = A(r)e^{i\omega t} \qquad F_r = \frac{B(r)}{f}e^{i\omega t}
\end{align}
The resulting equations for $A(r)$ and $B(r)$ are:
\begin{align}
\label{A}&\frac{d^2A}{dr^2}+\frac{n}{r}\frac{dA}{dr}+\left(-\frac{n}{r^2}-\frac{\alpha}{fr^2}-
\frac{\lambda^2}{f}+\frac{\omega^2}{f^2}\right)A=\left(\frac{2}{rf}-\frac{(n-1)b^{n-1}}{f^2r^n}\right)i\omega B \\
\label{B}&\frac{d^2B}{dr^2}+\frac{(n-2)}{r}\frac{dB}{dr}
+\left(-\frac{2(n-1)}{r^2}-\frac{\alpha}{fr^2}-\frac{\lambda^2}{f}+\frac{\omega^2}{f^2}\right)B=
-\frac{(n-1)b^{n-1}}{f^2r^n}i\omega A
\end{align}

In order to analyze the coupled equations (\ref{A}) and (\ref{B}), it is necessary to resort to the large $n$ limit, where $n$ is the number of dimensions of the sphere part of the metric. The various motivations for the large $n$ limit are discussed in \cite{emparan}. Here, we will only summarize the main steps of the method in \cite{emparan} and what we hope to achieve in our analysis of the coupled equations.

The function $f=\left(1-\frac{b^{n-1}}{r^{n-1}}\right)$ which appears in the background metric (\ref{giv1}) is an increasing function and $f(r) \to 1$ as $r \to \infty$. In the large $n$ limit, this function increases steeply from zero in the interval $b < r < b + \frac{b}{n}$ and is almost constant for $ r > \frac{b}{n}$. The appearance of distinct regions with a steep change in $f(r)$ in this limit is suited to the application of the method of matched asymptotic expansions.
First we define a near-horizon region and far region as follows:

\begin{align*}
\text{Near region} \quad &r-b \ll b \\
\text{Far region}\quad &r-b\gg\frac{b}{n-1}
\end{align*}

The definition of the near region is standard. The definition of the far region as done here is made possible by the large $n$ limit, in which $f(r)$ is almost constant in the far region. The two regions overlap in $\frac{b}{n-1} \ll r-b \ll b$.

The next step is to define a new coordinate $R=(\frac{r}{b})^{n-1}$. In term of this coordinate, the near and far regions are
\begin{align*}
\text{Near region} \quad & \ln R \ll n-1 \\
\text{Far region}\quad & \ln R \gg 1
\end{align*}

In the near-region approximation, $r$ can be written in terms of $R$ as
\begin{equation}\label{rR}
r \sim b \left[1+\frac{\ln R}{n-1}\right]
\end{equation}

We will now look at the coupled equations (\ref{A}) and (\ref{B}) in the near and far regions. The equations decouple in the large $n$ approximation in both these regions and can be solved. Then the far limit of the near region solution satisfying appropriate boundary conditions at the horizon and the near limit of an appropriate far region solution are compared in the overlap region to see if they can be matched. For discussions on stability, we need to investigate if there are normalizable solutions to the set of coupled equations that are regular at the horizon, with  $\omega = - i \Omega$ and $\Omega$ real (so that the solution grows in time). We will show that there are no such solutions, indicating the stability of the black string (brane) under vector perturbations. The same analysis can be applied for other choices of boundary conditions, such as those corresponding to quasinormal modes.

\subsection{The equations in the near region, large $n$ approximation}
We wish to analyze (\ref{A}) and (\ref{B}) in the near region. We also substitute $i\omega=\Omega$ in order to study black string (brane) stability. Among like terms in these equations, we only keep pieces which are of leading order in $n$. We assume $k_v, \lambda$ and $\omega$ to be at least of order $n$. We use the notation
$k_{v}^{2}/n^2=\hat{k_v}^2$, $\lambda^2/n^2=\hat{\lambda}^2$, $i\omega =\Omega$ and $\Omega^2/n^2 = \hat{\Omega}^2$.
To study solutions where they are of lower order in $n$, we can simply set them to zero.  As $\alpha =  k_{v}^{2}-(n-1)$, we replace it by $k_{v}^{2}$ for large $n$.  We then rewrite the equations in terms of the variable $R=(\frac{r}{b})^{n-1}$. To write functions of $r$ in terms of the variable $R$ in the equations, we use the approximate relation (\ref{rR}) which is valid in the near region. We also implicitly assume an expansion of $A$ and $B$ as
\begin{eqnarray}\label{n-exp-AB}
A=\sum_{i\geq 0}\frac{A_i}{n^i} \qquad B=\sum_{i\geq 0}\frac{B_i}{n^i}
\end{eqnarray}

Thus, in the near region, large $n$ approximation, the equations obeyed by $A$ and $B$ are
\begin{align}\label{nearAB}
\frac{d^2A}{dR^2}+\frac{2}{R}\frac{dA}{dR} - \left[\frac{\hat{k}_v^2}{R(R-1)} +
\frac{\hat{\lambda}^2b^2}{R(R-1)}+\frac{\hat{\Omega}^2 b^2}{(R-1)^2}\right]A  ~~~~~~~~~~~~~~~~~~~~~~~& & \nonumber \\= -\frac{\hat{\Omega}b}{R(R-1)^2}B +
\frac{2 \hat{\Omega}b}{n R (R-1)}B .& & \nonumber \\
\frac{d^2B}{dR^2}+\frac{2}{R}\frac{dB}{dR}-\left[\frac{\hat{k}_v^2}{R(R-1)}+
\frac{\hat{\lambda}^2b^2}{R(R-1)}+\frac{\hat{\Omega}^2 b^2}{(R-1)^2}\right]B = -\frac{\hat{\Omega}b}{R(R-1)^2}A. & &
\end{align}

We notice that the right-hand side of the first equation in (\ref{nearAB}) contains a term of the form $\frac{2 \hat{\Omega}b}{n R (R-1)}B $, which seems sub-leading in $n$ in comparison to a similar term $\frac{\hat{k}_v^2 + \hat{\lambda}^2b^2 }{ R (R-1)}A $ on the left-hand side. This statement is true if the leading order behaviour in $n$ of $A$ and $B$ in (\ref{n-exp-AB}) is similar. Indeed, this is the likely scenario in such systems of coupled equations. In such a case, this term on the right can be dropped. If on the other hand, $A$ is sub-leading in $n$ in comparison with $B$, the term must be retained. We will analyze both cases (while discussing the second case, we will also discuss the possibility of $B$ being sub-leading in comparison with $A$).
\vskip 0.3cm
{\bf Case 1}: Leading order behaviour in $n$ of $A$ and $B$ is similar.
\vskip 0.3cm
We can take $A = A_0 + A_{1}/n + ...$ and $B = B_0 + B_{1}/n + ...$, where $A_0, B_0 \neq 0$. In what follows, we will drop the subscripts of the leading terms in $n$ in $A$ and $B$.
In this case, in the large $n$ limit, (\ref{nearAB}) reduces to
\begin{subequations}
\begin{align}\label{nearABCase1}
\frac{d^2A}{dR^2}+\frac{2}{R}\frac{dA}{dR} - \left[\frac{\hat{k}_v^2}{R(R-1)} +
\frac{\hat{\lambda}^2b^2}{R(R-1)}+\frac{\hat{\Omega}^2 b^2}{(R-1)^2}\right]A  = -\frac{\hat{\Omega}b}{R(R-1)^2}B .  \\
\frac{d^2B}{dR^2}+\frac{2}{R}\frac{dB}{dR}-\left[\frac{\hat{k}_v^2}{R(R-1)}+
\frac{\hat{\lambda}^2b^2}{R(R-1)}+\frac{\hat{\Omega}^2 b^2}{(R-1)^2}\right]B = -\frac{\hat{\Omega}b}{R(R-1)^2}A.
\end{align}
\end{subequations}
It is clear from the form of (\ref{nearABCase1}) that a simple sum and difference of the two equations decouples them. We define
\begin{align*}
\xi=(R-1)^{-\hat{\Omega}b}(A+B) \qquad \zeta=(R-1)^{\hat{\Omega}b}(A-B)
\end{align*}

The equation obeyed by $\xi$ is
\begin{equation}
R(1-R)\frac{d^2\xi}{dR^2}+\left[2-(2\hat{\Omega}b+2)R\right]\frac{d\xi}{dR}-[\hat{\Omega}b-(\hat{k}_v^2+\hat{\lambda}^2b^2)]\xi=0.
\label{xieqn}
\end{equation}
This is an hypergeometric equation whose solutions, for $2\hat{\Omega}b$ not an integer are:
\begin{equation}
\xi=C_1F(p,q,2\hat{\Omega}b;1-R)+C_2(R-1)^{1-2\hat{\Omega}b}F(2-p,2-q,2-2\hat{\Omega}b;1-R);
\label{xisoln}
\end{equation}
where
\begin{align*}
p=\frac{1}{2}\left[1+2\hat{\Omega}b+\sqrt{1+4\hat{\Omega}^2b^2+4(\hat{k}_v^2+\hat{\lambda}^2b^2)}\right] \\ q=\frac{1}{2}\left[1+2\hat{\Omega}b-\sqrt{1+4\hat{\Omega}^2b^2+4(\hat{k}_v^2+\hat{\lambda}^2b^2)}\right]
\end{align*}
Physical considerations require that $A,B$ to be, at the very least, finite at the horizon.
For $\hat \Omega b > 1$, this implies $C_2 =0$. For $\hat \Omega b < 1$, both linearly independent solutions for $(A+B)$ approach zero as $R \to 1$. However, in fact, we need finiteness of the perturbation variables $F_t$ and $F_r$ at the horizon, which are related to $A,B$ by (\ref{modalF}). This requires $\hat \Omega b > 1$. Henceforth, we shall assume this.
We will impose the boundary condition $C_2 = 0$. The solution for $(A+B)$ is
\begin{equation}
(A+B)=(R-1)^{\hat{\Omega}b}C_1F(p,q,2\hat{\Omega}b;1-R)
\end{equation}
The equation and general solution for $\zeta$ can be obtained by replacing $\hat{\Omega}b$ by $-\hat{\Omega}b$ in (\ref{xieqn}) and (\ref{xisoln}) respectively. Since $(A-B) =(R-1)^{- \hat{\Omega}b}\zeta $, the solution $(A-B)$ that is regular at the horizon is now given by $C_1 = 0$.

For $\hat{\Omega}b = N$ a positive integer, and $p \neq 1,2,...., 2N-1$, the general solution for $\xi$ is now
$$\xi=C_1F(p,q,2\hat{\Omega}b;1-R)+C_2 \ln (R-1) F(p, q,2\hat{\Omega}b;1-R).$$
Finiteness of the perturbation at the horizon implies $C_2 = 0$. If $p$ is one of the integers $1,2,...., 2N-1$, then the general solution for $\xi$ is given by (\ref{xisoln}). We will not discuss these cases further as the finite solution in all cases is the same.
\vskip 0.3cm
{\bf Case 2}: $A$ is sub-leading in $n$ in comparison to $B$ (or vice-versa).
\vskip 0.3cm
If $A$ is sub-leading in $n$, then in the expansion (\ref{n-exp-AB}), where $A = A_0 + A_{1}/n +...$,
 and $B = B_0 + B_{1}/n +...)$, we set $A_0 = 0$. The equations (\ref{nearAB}) then imply $B_0 = 0$ which brings us back
to {\bf Case 1}. A similar analysis follows when $B_0 = 0$.

\subsection{The far region in the large $n$ approximation}
The far region is defined by $r\gg b+\frac{b}{n}$. Therefore, in this limit $f\rightarrow 1$ as $(b^{n-1}/r^{n-1}) \sim e^{-n\ln r}$ is a small quantity for large $n$ and large $r$. We can neglect terms that have $f'$ (or $f''$) in (\ref{A}) and (\ref{B}) because they fall off at least as $b^{n-1}/r^{n}$ and are negligible compared to other terms that fall off as $1/r^2$. We then use the large $n$ approximation to retain only the leading $n$ parts in like terms. To consider the most general case, we have assumed $k_{v}^2 , \Omega^2 = - \omega^2$ and $\lambda^2$ are of order $n^2$. To consider the case when these quantities are of lower order in $n$,  we can take them to zero in our final answer.

We observe, that, for example, the term on the right-hand side of (\ref{B}) is, in this limit,  $-\frac{(n-1)b^{n-1}}{ r^n}\Omega A$. Decaying terms we consider in this approximation on the left-hand side are of the form $\frac{1}{r^2}B $. For the two to be comparable, we need at least $A \sim r^{n-2}B$. If this were true, we could neglect the right-hand side of the equation for $A$ (\ref{A}) in this limit. If this were not true, then the right-hand side of the equation for $B$ (\ref{B}) can be neglected in this limit. In either case, one of the equations will have the right-hand side zero - one could solve this and substitute the solution in the other equation as a source term. We also note that in either situation, we additionally require normalizability of both sets of perturbations. We will assume that the right-hand side of equation (\ref{B}) can be neglected. The other case - neglecting the right-hand side of the equation (\ref{A}) is almost identical computationally. This is due to the fact that in the large $n$ approximation, noting that $\lambda^2$, for example, is at least of order $n^2$, the left-hand side of (\ref{B}) is identical to that of (\ref{A}) with the replacement of $A$ by $B$. The only difference between the two cases is in the type of source term in each of the equations.

Neglecting the right-hand side of (\ref{B}), we have, in the large $n$ far region,
\begin{align}\label{Brhs}
&\frac{d^2A}{dr^2}+\frac{n}{r}\frac{dA}{dr}+\left(-\frac{k_v^2}{r^2}-\lambda^2-\Omega^2\right)A=\left(\frac{2}{r}\right)\Omega B \\
&\frac{d^2B}{dr^2}+\frac{n}{r}\frac{dB}{dr}+\left(-\frac{k_v^2}{r^2}-\lambda^2-\Omega^2\right)B=0
\end{align}
The general solution for $B$ is given in terms of modified Bessel functions of order $\nu = \sqrt{\frac{(n-1)^2}{4}+k_v^2}$ as
\begin{equation}
B=r^{\frac{1-n}{2}}[D_1 I_\nu(\sqrt{\lambda^2+\Omega^2} r)+D_2K_\nu(\sqrt{\lambda^2+\Omega^2} r)]
\label{farB}
\end{equation}

We note that the large $n$ limit implies the limit of large order and large argument for the modified Bessel functions, due to $\nu \sim O(n)$ and
$\sqrt{\lambda^2+\Omega^2} \sim O(n)$. Rewriting $\sqrt{\lambda^2+\Omega^2} r = \nu z$ where $z = \frac{ \sqrt{\lambda^2+\Omega^2}}{\nu} r$, we use standard expansions for large order and large argument for the modified Bessel functions. $I_\nu(\nu z) \sim e^{\nu z} \to \infty$ as $r \to \infty $, whereas $K_\nu(\nu z) \sim e^{- \nu z} \to 0$. Normalizability thus implies in (\ref{farB}) that $D_1 = 0$.

We now need to match the solutions in the near region and the far region in the overlap region $\frac{b}{n-1} \ll r-b \ll b$. To do this, first we use the asymptotic expansion for large order and large argument of the modified Bessel functions in $B$ to obtain
\begin{equation}
B = D_2 r^{\frac{1-n}{2}} \sqrt{\frac{\pi}{2 \nu}} e^{- \nu \sqrt{1 + z^2}} \lb \frac{1 + \sqrt{1 + z^2}}{z}\rb^{\nu} (1+z^2 )^{-\frac{1}{4}} \ltb 1 + O(\frac{1}{n}) \rtb.
\label{asyB}
\end{equation}
The form of (\ref{asyB}) in the overlap region suitable for matching is given by changing variables from $r$ to $R$ in (\ref{asyB}) using the approximate formula (\ref{rR}) valid in the overlap region. In the large $n$ approximation, this is
\begin{equation}
B=D_1 R^{-\frac{1}{2}-\frac{\sqrt{1+4\hat{\Omega}^2b^2+4(\hat{k}_v^2+\hat{\lambda}^2b^2)}}{2}}
\label{asyBoverlap}
\end{equation}
The relevant asymptotic expansion of the modified Bessel function used here is given in Appendix D.

We need to substitute the expression for $B$ from (\ref{asyB}) and (\ref{asyBoverlap}) in the right-hand side of the first equation in (\ref{Brhs}) to get $A$. As can be checked (for details of handling source terms, we refer to Appendix B), $A$ has the same behaviour as $B$ as $r \to \infty$. Further, in the overlap region, we have the same power law behaviour,
$A = (const.) R^{-\frac{1}{2}-\frac{\sqrt{1+4\hat{\Omega}^2b^2+4(\hat{k}_v^2+\hat{\lambda}^2b^2)}}{2}}$. Thus $A+B$ and $A-B$ have the same behaviour.

\subsection*{Matching of Solutions}
For matching, we need to extend the near region solution to overlap region. In order to do this, we use the transformation properties of hypergeometric functions relating functions of argument $(1-R)$ to those of argument $1/R$. We will first consider the variable $A+B$ in sub-section III.1.
\begin{align*}
A+B &=(R-1)^{\hat{\Omega}b}C_1F(p,q,2\hat{\Omega}b;1-R)\\
  &=(R-1)^{\hat{\Omega}b}C_1\big[\tilde{c_1}R^{-p}F(p,p-1,p-q+1;1/R)\\
  &\hspace*{3 cm}+\tilde{c_2}R^{-q}F(q,q-1,q-p+1;1/R)\big];
\end{align*}
where constants $\tilde{c_1},\tilde{c_2}$ depend on $p,q$ .
To extend the solution to the overlap region, we put $(R-1)\approx R$ as we are sufficiently far from horizon and take the limit $R \rightarrow \infty$ in hypergeometric function.
After this extension, explicitly putting the values of $\tilde{c_1}$ and $\tilde{c_2}$, we get

\begin{align}\label{A+Boverlap}
A+B=C_1\frac{\Gamma(p+q-c+1)\Gamma(q-p)}{\Gamma(q)\Gamma(q-c+1)}R^{-\frac{1}{2}-\frac{\sqrt{1+4\hat{\Omega}^2 b^2+4(\hat{k}_v^2+\hat{\lambda}^2b^2)}}{2}}+\nn[0.2 cm]
C_1\frac{\Gamma(p+q-c+1)\Gamma(p-q)}{\Gamma(p)\Gamma(p-c+1)}R^{-\frac{1}{2}+\frac{\sqrt{1+4\hat{\Omega}^2b^2+4(\hat{k}_v^2+\hat{\lambda}^2b^2)}}{2}}
\end{align}

Here $c$ is a constant from the original hypergeometric equations and $c=2$.
As we are looking for stable solutions for $\lambda, \Omega$ positive, for such values
\begin{equation}
\frac{1}{2}<\frac{\sqrt{1+4\hat{\Omega}^2b^2+4(\hat{k}_v^2+\hat{\lambda}^2b^2)}}{2} ;
\end{equation}
hence we get a growing solution in $R$ with pieces $R^{-1/2 + d}$ and $R^{-1/2 - d}$ where $d = \frac{\sqrt{1+4\hat{\Omega}^2 b^2+ 4(\hat{k}_v^2+\hat{\lambda}^2b^2 )}}{2}$. Note that this solution will always have the growing piece  $R^{-1/2 + d}$ since its coefficient does not vanish for any $\hat{\Omega} \geq 0$. This can be seen as follows:

The coefficient of the growing piece is
$$\frac{\Gamma(p+q-c+1)\Gamma(p-q)}{\Gamma(p)\Gamma(p-c+1)},$$
and the Gamma function is non-zero. The coefficient can go to zero only at the poles of Gamma functions in the denominator. This can only occur when either $p$ or $p-c+1$ is a non-positive integer. These two quantities are;
\begin{align*}
&p=\frac{1}{2}\left[1+2\hat{\Omega}b+\sqrt{1+4\hat{\Omega}^2b^2+4(\hat{k}_v^2+\hat{\lambda}^2b^2)}\right] \\
&p-c+1=-\frac{1}{2}+\hat{\Omega}b+\frac{\sqrt{1+4\hat{\Omega}^2b^2+4(\hat{k}_v^2+\hat{\lambda}^2b^2)}}{2}.
\end{align*}

It is easy to see that for our situation when $\hat{\Omega} \geq 0$ and $\hat{k}_v >0$ these quantities can never be non-positive integers. Thus the solution from the near region will always have the growing piece.
But the normalizable solution in the far region is
\begin{equation}
A+B= (Const.) R^{-\frac{1}{2}-\frac{\sqrt{1+4\hat{\Omega}^2b^2+4(\hat{k}_v^2+\hat{\lambda}^2b^2)}}{2}};
\end{equation}
which is a decaying solution. Therefore we cannot match the solutions from near and far region. There are no unstable solutions for $\lambda, \Omega$ positive. A similar statement holds for $(A-B)$. Its expansion in the overlap region is similar to (\ref{A+Boverlap}) in that it has both pieces $R^{-1/2 + d}$ and $R^{-1/2 - d}$. The normalizable solution in the far region extended to the overlap region has only the decaying term $R^{-1/2 - d}$. Hence a match is not possible.

So far, we have considered $k_v, \Omega, \lambda \sim O(n)$ at least. We could consider $\Omega, \lambda$ of lower order by setting them to be zero. The match is still not possible unless we additionally set $k_v = 0$. Since the IK variables are not defined for $k_v = 0$, we cannot set it to be zero. In particular, in the static limit, with $\Omega = 0$, we have no instability.

Thus we conclude that for $\hat \Omega $ real and non-negative, there are no normalizable solutions to the vector perturbation equations, and the black string/brane is stable.

By considering $\omega = - i \Omega$ and looking for oscillatory solutions, we find that in the asymptotic region, the modified Bessel functions are replaced by Bessel functions. Both the linearly independent Bessel functions have similar behaviour asymptotically, and can be considered. Hence the solution from the far region will be any arbitrary linear combination of the pieces $R^{-1/2 + d}$ and $R^{-1/2 - d}$ with the replacement $\omega = - i \Omega$ in $d$. It can be matched to a solution from the near-region obeying suitable boundary conditions at the horizon. This can be used to construct approximate solutions to the perturbation equations. One can also consider boundary conditions suitable for quasinormal modes, such as ingoing at the horizon and outgoing at infinity, and use the matching procedure to evaluate the quasinormal modes.

\section{Non-spherically symmetric scalar perturbations}
The IK variables defined for scalar perturbations, $F_{ab}$ and $F$ are (\ref{Fab})

\begin{align*}
F_{ab}=f_{ab}+D_aX_b+D_bX_a\\
F=H_L+\frac{H_T}{n}-\frac{D^ar}{r}X_a.
\end{align*}

Our goal is to write eigenvalue equations
$2\delta G_{\mu\nu}=-\lambda^2 h_{\mu\nu}$ in terms of the IK variables to make them more tractable, where $\delta G_{\mu\nu}$ is the linearized Einstein tensor of the Schwarzschild-Tangherlini black hole metric. This involves a lengthy computation for which we outline the steps. We first use \cite{iks} to express $2\delta G_{\mu\nu}$ in terms of $F_{ab}$ and $F$. We note that the right hand side of the eigenvalue equations is not expressed in terms of either $F_{ab}$ of $F$.  There are four IK variables $(F_{rr},F_{rt},F_{tt},F)$. To simplify the equations further, following IK \cite{vac}, we construct three functions from $F_{ab}, F$;
\begin{align}\label{Sca-WYZ}
& W=r^{n-2}(F^t_t -2F) \qquad Y=r^{n-2}(F^r_r -2F) \qquad Z=r^{n-2}F^r_t.
\end{align}

The goal is to take appropriate combinations of the eigenvalue equations for $\delta G_{\mu\nu}$ so that they can be expressed entirely in terms of just three perturbation variables $W,Y$ and $Z$.
To accomplish this, we can invert the relations (\ref{Sca-WYZ}) to obtain $F_a^a$ and $F$ in terms of $W,Y$ and $Z$. For this, we need the traceless part of the equation $2\delta G_{ij}=-\lambda^2h_{ij}$ (written partly in terms of the new variables):
\begin{equation}\label{Sca-WYF}
W+Y+2nF=2\lambda^2\frac{r^2}{k^2}H_T
\end{equation}
We can write $F$ in terms of $W$, $Y$ and $H_T$ using this relation. Subsequently using (\ref{Sca-WYF}) and (\ref{Sca-WYZ}) we get:
\begin{subequations}\label{Sca-F_to_WY}
\begin{align}
&F=-\frac{W+Y}{2n r^{n-2}}+\frac{\lambda^2}{nk^2}(r^2H_T) \\ &F^r_t=\frac{Z}{r^{n-2}} \\
&F^t_t=\frac{W(n-1)-Y}{2n r^{n-2}}+\frac{2\lambda^2}{nk^2}(r^2H_T) \\ &F^t_t=\frac{Y(n-1)-W}{2n r^{n-2}}+\frac{2\lambda^2}{nk^2}(r^2H_T)
\end{align}
\end{subequations}
Our choice of variables is motivated by those of Ishibashi and Kodama who studied the linearized Einstein equation --- their variables therefore correspond to (\ref{Sca-WYZ}) with $\lambda$ is zero. Hence their expressions of $F_a^a,F$ in terms of $W,Y$ and $Z$ do not have the factors of $H_T$ that ours have.

Substituting our new variables in the eigenvalue equations, we obtain six equations (\ref{f_t})-(\ref{f_tt}) which are given in Appendix C. Due to the $H_T$ factors in (\ref{Sca-F_to_WY}), these equations have terms containing derivatives of $H_T$ in addition to components of $h_{\mu\nu}$.

Our goal is to get the final equations completely in terms of $W,Y$ and $Z$ by taking suitable combinations of the eigenvalue equations, in analogy with the work of Ishibashi and Kodama, and despite the extra $H_T$ factors present in
our expressions. After a lengthy calculation, we have succeeded in obtaining the final equations completely in terms of these variables, and all the $H_T$ factors cancel out. These equations and the relevant details are given in (\ref{Z})-(\ref{W}) in Appendix C. The important fact for us in the subsequent analysis is that
the scalar perturbation equations can be reduced to three coupled second order partial differential equations for
$W$, $Y$ and $Z$. Our later computations are made simpler by the further change of variables:
\begin{align}\label{Sca-def-sifi}
\hat{\psi}= \frac{f^{1/2}}{r^{(n-4)/2}} W \qquad \hat{\phi} = \frac{f^{1/2}}{r^{n/2}} Y \qquad \hat{\eta} = \frac{1}{r^{(n-2)/2}f^{1/2}}Z ;
\end{align}
where $f=\left(1-\frac{b^{n-1}}{r^{n-1}}\right)$.
We can assume a time dependence of the form $\hat{\psi}(r,t)=e^{i\omega t}\psi(r)$ for all three variables. Finally, the
three coupled perturbation equations are:
\begin{align}\label{psi}
&-\frac{d^2\psi}{dr^2}+\bigg[\frac{n^3-2n^2+8n-8}{4nr^2}+\frac{f'^2}{4f^2}-\frac{(n^2+2n-4)}{2n}\frac{f'}{fr}-\frac{f''}{2f} \nonumber\\
&-\frac{2(n-1)}{nr^2f}+\frac{k^2}{fr^2}+\frac{\lambda^2}{f}-\frac{\omega^2}{f^2}\bigg]\psi = \nonumber\\
&\left[\frac{4}{f}-\frac{2f'r}{f^2}\right](i\omega)  \eta+\left[\frac{2(n-1)}{nf}+\frac{2}{n}-\frac{n+2}{n}\frac{rf'}{f}-\frac{r^2 f''}{f}+\frac{f'^2r^2}{2f^2}\right]\phi
\end{align}
\begin{align}
\label{phi}&-\frac{d^2\phi}{dr^2}+\bigg[\frac{n^3-2n^2+8n-8}{4nr^2}+\frac{f'^2}{4f^2}-
\frac{(n^2+2n-4)}{2n}\frac{f'}{fr}-\frac{f''}{2f} \nonumber \\
&-\frac{2(n-1)}{nr^2f}+\frac{k^2}{fr^2}+\frac{\lambda^2}{f} -\frac{\omega^2}{f^2}\bigg]\phi = \nonumber \\
&\frac{2f'}{f^2r}\eta(i\omega) +\left[\frac{2(n-1)}{nr^4f}-\frac{2(n-1)}{nr^4}-\frac{2-n}{nr^3}\frac{f'}{f}-\frac{f''}{r^2f}+\frac{f'^2}{2f^2r^2}\right]\psi
\end{align}
\begin{align}
\label{eta}&-\frac{d^2\eta}{dr^2}+\left[\frac{n^2-2n}{4r^2}-\frac{(n+2)f'}{2rf}+\frac{3f'^2}{4f^2}-\frac{3f''}{2f} +\frac{k^2}{fr^2}-\frac{\omega^2}{f^2}+\frac{\lambda^2}{f}\right]\eta \nonumber \\
&\hspace*{5 cm} = \left[\frac{f'}{f}-\frac{2}{r}\right]
\frac{r(i\omega)}{f}\phi-\frac{f'}{f^2}\frac{(i\omega)}{r}\psi
\end{align}

Although we have described the details in Appendix C, we will state the eigenvalue equations that have been used to
obtain these coupled equations.
The equation for $\phi$ is obtained by combining the eigenvalue equations corresponding to $\delta G_{rr}$, trace of $\delta G_{ij}$ , $\delta G_{ri}$ and $\delta G_{ti}$. Similarly, the equation for $\psi$ is obtained by taking suitable combinations of equations corresponding to $\delta G_{rr}$, trace of $\delta G_{ij}$ , $\delta G_{ri}$ and $\delta G_{ti}$. The equation for $\eta$ is obtained by combining $\delta G_{rt}$, $\delta G_{ri}$ and $\delta G_{ti}$.

The equations (\ref{psi})---(\ref{eta}) cannot be solved analytically. In order to solve them, we have to take the large $n$ limit of the equations. The equations do not decouple,
but as we did for the vector case, we analyze the large $n$ limit of (\ref{psi})---(\ref{eta}) in the near-horizon and
far regions. We then investigate using matched asymptotic expansions, if the two solutions match in the overlap region.

\subsection{Far region}

The far region is defined, as before, by $r\gg b+\frac{b}{n}$. In this limit, we will take $f\rightarrow 1$  and use the
large $n$ limit as well, to neglect
terms that have $f', f''$ in the equations for $\psi$, $\phi$ and $\eta$. Further, we retain only leading order in $n$ pieces
in like terms. We have assumed $k^2, \omega^2$ and $\lambda^2$ are at least of order $n^2$ as it is the most general case.
We can recover cases where these quantities are of lesser order by putting them to zero in our final answer (with some minor
modifications), except for $k=0$ (which would correspond to the s-mode).\\
In the far limit, the equations (\ref{phi})---(\ref{eta}) are (replacing $\Omega = i \omega$ since we want to do a stability
analysis):

\begin{align}\label{far-sifi1}
&-\frac{d^2\psi}{dr^2}+\left[\frac{n^2}{4r^2}+\frac{k^2}{r^2}+\lambda^2+\Omega^2\right]\psi=2\phi+4\Omega\eta
\\[0.3 cm]
\label{far-sifi2}&
-\frac{d^2\phi}{dr^2}+\left[\frac{n^2}{4r^2}+\frac{k^2}{r^2}+\lambda^2+\Omega^2\right]\phi=
\frac{-f''}{f r^2}\psi+\frac{2f'}{f^2 r}\Omega \eta\\[0.3 cm]
\label{far-sifi3}&
-\frac{d^2\eta}{dr^2}+\left[\frac{n^2}{4r^2}+\frac{k^2}{r^2}+\lambda^2+\Omega^2\right]\eta=
-2\Omega \phi-\frac{f'}{f^2 r} \Omega \psi
\end{align}
We first consider the $\phi$ equation. We have listed terms on the left-hand side of the type $(\phi/r^2)$.
We note that terms on the right-hand side are of the type $ (\psi/r^{n+3}), (\eta/r^{n+1})$. For any of these terms to
be considered, for example, $(\eta/r^{n+1})$, the magnitude of $\eta$ must be at least $\eta \sim \phi r^{n-1}$. The same
argument also holds for $\psi$ - its magnitude must be at least $\psi \sim \phi r^{n+1}$ for the term proportional
to it on the right to be considered. As we will discuss later, there is no solution to the equations (\ref{psi})---(\ref{eta})
in the large $n$ limit, corresponding to this situation.
Hence, in what follows, we will assume that $\eta$ and $\psi$ are comparable in
in magnitude to $\phi$ in the large $n$ limit so that the right-hand side
of (\ref{far-sifi2}) can be neglected.

We can solve for $\phi$ and subsequently solve equation for $\psi$ and $\eta$.
\begin{equation}\label{far-phi}
-\frac{d^2\phi}{dr^2}+\left[\frac{n^2}{4r^2}+\frac{k^2}{r^2}+\lambda^2+\Omega^2\right]\phi=0
\end{equation}
Let $\nu = \sqrt{\frac{n^2+1}{4}+k^2}$. The general solution of (\ref{far-phi}) is given in terms of modified Bessel functions
as:
\begin{equation}\label{far-phi-gsol}
\phi=D_1 \sqrt{r}I_\nu(\sqrt{\lambda^2+\Omega^2} r)+D_2 \sqrt{r}K_\nu(\sqrt{\lambda^2+\Omega^2} r)
\end{equation}

The order $\nu$ of the modified Bessel functions is proportional to $n$. As we have $\lambda,\Omega$ of order $n$, the
large $n$ limit implies the large order and argument limit of the modified Bessel functions.
In this limit,  $I_\nu(\nu z) \sim e^{\nu z}$ is a growing solution and for normalizability, we need to choose $D_1=0$.
Hence
\begin{equation}\label{far-phi-sol}
\phi=D_2\sqrt{r}K_\nu(\sqrt{\lambda^2+\Omega^2} r).
\end{equation}

As in the vector case, we need to find the expansion of this solution in the overlap region next.
We use the large argument and order expansion of $K_\nu(\sqrt{\lambda^2+\Omega^2} r)$
and retain terms to leading order in $n$. We then change coordinates from $r$ to $R$ using (\ref{rR}) which
is valid in the overlap region. The outline of this calculation is given in Appendix D.
In terms of $R$, the leading order far solution for $\phi$ in the overlap region becomes
\begin{equation}\label{farphi}
\phi= D_0 R^{-\frac{\sqrt{1+4(\hat{k}^2+\hat{\lambda}^2 b^2)+4\hat{\Omega}^2b^2}}{2}}
\end{equation}
We henceforth denote scaled $k$, $\lambda$ as $k^2/n^2=\hat{k}^2$ and  $\lambda^2/n^2=\hat{\lambda}^2$.
To solve the equations for $\psi$ and $\eta$, we will use (\ref{far-phi-sol}). The $\eta$ equation (\ref{far-sifi3}) is
tackled first. In this equation, the term proportional to $\psi$ on the right can be neglected, due to the $f'$ term.
$-2 \Omega \phi$, with $\phi$ given by (\ref{far-phi-sol}) appears as a source on the right in this equation.
We can find a particular solution to this equation by the method of variation of parameters exactly as we did in the vector
case (detailed in Appendix B). The computation is similar, and the relevant facts are: the
particular solution will decay exponentially as $r \to \infty$, and will have the power law behaviour
$\eta = (const.) R^{-\frac{\sqrt{1+4(\hat{k}^2+\hat{\lambda}^2 b^2)+4\hat{\Omega}^2b^2}}{2}}$ in the overlap region.
A similar statement can be made for $\psi$.

\subsection{Near region}

For the near region behaviour, it is convenient to analyze equations (\ref{psi})---(\ref{eta}) in the
$R$ variable given in the near region by (\ref{rR}).
We also expand $\phi$, $\psi$ and $\eta$ in terms of $n$ as
\begin{align}\label{near-sifi-n}
\psi=\sum_{i\geq0}\frac{\psi_i(R)}{n^i} \qquad \phi=\sum_{i\geq0}\frac{\phi_i(R)}{n^i} \qquad \eta=\sum_{i\geq0}\frac{\eta_i(R)}{n^i}
\end{align}

Taking the large $n$ limit of (\ref{psi})---(\ref{eta}), we obtain:
\begin{align}
\label{near-psi}&\frac{d^2\psi}{dR^2}+\frac{1}{R}\frac{d\psi}{dR}-\left[\frac{1}{4R^2}+\frac{1}{4(R-1)^2R^2}-\frac{1}{nR^2(R-1)}+\frac{\hat{k^2}+\hat{\lambda}^2b^2}{R(R-1)}+\frac{\hat{\Omega}^2b^2}{(R-1)^2}\right]\psi=\nonumber \\
&\hspace*{2.2 cm}-\left[\frac{2}{n^2(R-1)R}+\frac{2}{n^3R^2}+\frac{1}{R^2(R-1)}+\frac{1}{2R^2(R-1)^2}\right]\phi b^2 \nn
&\hspace*{2.2 cm}-\left[\frac{4}{nR(R-1)}-\frac{2}{R(R-1)^2}\right]\hat{\Omega}b^2\eta
\end{align}
\begin{align}
\label{near-phi}&\frac{d^2\phi}{dR^2}+\frac{1}{R}\frac{d\phi}{dR}-\left[\frac{1}{4R^2}+\frac{1}{4(R-1)^2R^2}-\frac{1}{nR^2(R-1)}+\frac{\hat{k^2}+\hat{\lambda}^2b^2}{R(R-1)}+\frac{\hat{\Omega}^2b^2}{(R-1)^2}\right]\phi=\nonumber \\
&-\left[\frac{2}{n^2(R-1)R}-\frac{2}{n^2R^2}+\frac{1}{R^2(R-1)}+\frac{1}{2R^2(R-1)^2}\right]\frac{\psi}{b^2}-\left[\frac{2}{R(R-1)^2}\right]\hat{\Omega}b^2\eta
\end{align}
\begin{align}
\label{near-eta}&\frac{d^2\eta}{dR^2}+\frac{1}{R}\frac{d\eta}{dR}-\left[\frac{1}{4R^2}+\frac{3}{4(R-1)^2R^2}+\frac{1}{R^2(R-1)}+\frac{\hat{k^2}+\hat{\lambda}^2b^2}{R(R-1)}+\frac{\hat{\Omega}^2b^2}{(R-1)^2}\right]\eta=\nonumber \\
&\hspace*{2.2 cm}-\left[\frac{1}{(R-1)^2R}-\frac{2}{n(R-1)R}\right]\hat{\Omega}\phi b^2+\frac{\hat{\Omega}}{R(R-1)^2}\psi
\end{align}

Our notation is: $k^2/n^2=\hat{k}^2$, $\lambda^2/n^2=\hat{\lambda}^2$, $i\omega =\Omega$ and $\Omega^2/n^2 = \hat{\Omega}^2$.

We have implicitly used the expansion (\ref{near-sifi-n}) in these equations. In fact, we are solving for the
leading term in the expansion of $\phi$, $\psi$ and $\eta$ in the large $n$ limit.
From the equations (\ref{near-psi})---(\ref{near-eta}), it is clear that the leading terms $\phi_0$, $\psi_0$ and $\eta_0$
in the expansion (\ref{near-sifi-n}) all have to be non-zero. Hence we will not use the subscripts in the following equations.

Taking linear combinations of $\phi$ and $\psi$ simplifies the system to a great extent.
Denote $\eta b^2=\tilde{\eta} \text{ and } \phi b^2=\tilde{\phi}$ , define
\begin{equation}\label{near-HG}
H=\psi+\tilde{\phi} \text{ and }  G=\psi-\tilde{\phi}.
\end{equation}

The equation for $H$ decouples as $\eta$ terms cancel out.
\begin{equation}\label{near-H-eqn}
\frac{d^2H}{dR^2}+\frac{1}{R}\frac{dH}{dR}-
\left[\frac{1}{4R^2}-\frac{1}{4R^2(R-1)^2}-\frac{1}{R^2(R-1)}+\frac{(\hat{k}^2+\hat{\lambda}^2 b^2)}{R(R-1)}+\frac{\hat{\Omega}^2b^2}{(R-1)^2}\right]H=0
\end{equation}

This equation can be written as an hypergeometric equation with regular singular
points at $0,1$ and $\infty$ by making the ansatz $H=R(R-1)^{\frac{1}{2}+\hat{\Omega}b}M$

\begin{equation}\label{near-M-eqn}
R(1-R)\frac{d^2M}{dR^2}+[3-(4+2\hat{\Omega} b)R]\frac{dM}{dR}+[(\hat{k}^2+\hat{\lambda}^2 b^2)-3\hat{\Omega}b-2]M=0
\end{equation}

If $1+ 2 \hat \Omega b \neq m$ where $m$ is a positive integer, the solution of this equation is of the form
\begin{equation}\label{near-M-sol}
M= C_1 F(p,q,1+2\hat{\Omega}b,1-R)+C_2(1-R)^{-2\hat{\Omega}b}F(3-p,3-q,1-2\hat{\Omega}b,1-R);
\end{equation}

where

\begin{align}\label{near-M-pq}
p=\frac{1}{2}\ltb 3+2\hat{\Omega}b-\sqrt{1+4(\hat{k}^2+\hat{\lambda}^2 b^2)+4\hat{\Omega}^2b^2}\rtb \nn
 q=\frac{1}{2}\ltb 3+2\hat{\Omega}b+\sqrt{1+4(\hat{k}^2+\hat{\lambda}^2 b^2)+4\hat{\Omega}^2b^2}\rtb
\end{align}

In this case, for $\hat{\Omega}b > \frac{1}{2}$, there is an unambiguous way to choose the behaviour of the solution
at the horizon. Finiteness of $H$ implies that in the general solution for $M$, we must set
$C_2 = 0$. For $\hat{\Omega}b \leq \frac{1}{2}$, both linearly independent solutions for $H$ are finite.
Thus there seems to be
some ambiguity in the choice of boundary condition at the horizon. As in the vector case, we argue that it is the original
perturbation variables that need to be finite at the horizon. Consulting (\ref{Sca-def-sifi}) for the definitions of the variables $\phi, \psi, \tilde{\eta}$
in terms of $W, Y, Z$, we see that finiteness of $W,Y,Z$ at the horizon implies that $C_2 = 0$ even in this case.

We thus set $C_2=0$.
Therefore in the near region,
\begin{equation}\label{near-H-sol}
H= C_1 R(R-1)^{\frac{1}{2}+\hat{\Omega}b}F(p,q,1,1-R)
\end{equation}

Even in the case $1+ 2 \hat \Omega b = m$, (\ref{near-H-sol}) is the appropriate solution for finiteness of the perturbation at the horizon.
For matching, we need to write the asymptotic expansion of the near region solution in the overlap region.
In order to do this, we use the standard transformation formula:
\begin{align}\label{near-H-hyp}
H&=R(R-1)^{\frac{1}{2}+\hat{\Omega}b}C_1F(p,q,1+2\hat{\Omega}b;1-R)\\
  &=R(R-1)^{\frac{1}{2}+\hat{\Omega}b}C_1\big[\tilde{c_1}R^{-p}F(p,p-2\hat{\Omega}b,p-q+1;1/R)+\nn
  &\qquad\qquad\qquad\qquad\tilde{c_2}R^{-q}F(q,q-2\hat{\Omega}b,q-p+1;1/R)\big];
\end{align}
where constants $\tilde{c_1},\tilde{c_2}$ depend on $p,q$.
The asymptotic expansion for $H$ in the overlap region is obtained by putting $(R-1)\approx R$
and evaluating the hypergeometric functions in the large $R$ approximation.

The asymptotic expansion of the near region solution for $H$ in the overlap region is
\begin{equation}\label{near-H-extn}
H=C_1\left[\tilde{c_1}R^{\frac{\sqrt{1+4\hat{\Omega}^2b^2+4(\hat{k}^2+\hat{\lambda}^2b^2)}}{2}}+\tilde{c_2}R^{-\frac{\sqrt{1+4\hat{\Omega}^2b^2+4(\hat{k}^2+\hat{\lambda}^2b^2)}}{2}}\right]
\end{equation}

 By the same reasoning as employed in vector case (\ref{A+Boverlap}), it can be argued that for $\hat{\Omega} \geq 0$ and $\hat{k}>0$, the coefficient of the growing piece $\tilde{c_1}$ is never zero for any value of $\hat{\Omega}$.

We would like to do a similar procedure for the other perturbation equations for $G$ and $\tilde{\eta}$.
The equations for $G$ and  $\tilde{\eta}$ in the near region are:
\begin{align}
\label{scalarG}&\frac{d^2G}{dR^2}+\frac{1}{R}\frac{dG}{dR}-\left[\frac{1}{4R^2}+\frac{3}{4R^2(R-1)^2}+\frac{1}{R^2(R-1)}+\frac{(\hat{k}^2+\hat{\lambda}^2 b^2)}{R(R-1)}+\frac{\hat{\Omega}^2b^2}{(R-1)^2}\right]G\nn
&\hspace{9 cm}=\frac{4\hat{\Omega}}{R(R-1)^2}\tilde{\eta}\\
\label{scalarEta}&\frac{d^2\tilde{\eta}}{dR^2}+\frac{1}{R}\frac{d\tilde{\eta}}{dR}-\left[\frac{1}{4R^2}+\frac{3}{4R^2(R-1)^2}+\frac{1}{R^2(R-1)}+\frac{(\hat{k}^2+\hat{\lambda}^2 b^2)}{R(R-1)}+\frac{\hat{\Omega}^2b^2}{(R-1)^2}\right]\tilde{\eta}\nn
&\hspace*{9 cm}=\frac{\hat{\Omega}b^2}{R(R-1)^2}G
\end{align}

Unfortunately they are coupled. We cannot solve these equations analytically in the near region as we did for $H$.
First, let us consider the special case when either one of $G$ or $\tilde{\eta}$ is zero. Since
the left-hand sides of both (\ref{scalarG}) and (\ref{scalarEta}) have the same differential operator, the two possibilities are computationally identical. Let us take the case $\tilde{\eta} = 0$. Then we obtain the following general solution for
$G$ (with $\tilde{\eta} = 0$):
\begin{align}
G = &D_1(R-1)^{\frac{1}{2} + \sqrt{1 + \hat \Omega^2 b^2}}  F( \tilde p, \tilde q, 1 + 2 \sqrt{1 + \hat \Omega^2 b^2}; 1-R)+ \nn
 &D_2 (R-1)^{\frac{1}{2} -  \sqrt{1 + \hat \Omega^2 b^2}} F(1-\tilde p, 1-\tilde q, 1 - 2 \sqrt{1 + \hat \Omega^2 b^2}; 1-R) .
\label{etazero}
\end{align}
Here, $$\tilde p = \frac{1}{2} + \sqrt{1 + \hat \Omega^2 b^2} + \frac{\sqrt{1+4\hat{\Omega}^2b^2+4(\hat{k}^2+\hat{\lambda}^2 b^2)}}{2}$$ and $$\tilde q = \frac{1}{2} + \sqrt{1 + \hat \Omega^2 b^2} - \frac{\sqrt{1+4\hat{\Omega}^2b^2+4(\hat{k}^2+\hat{\lambda}^2b^2)}}{2}.$$

Finiteness at the horizon implies we must set $D_2 = 0$.

This solution can be written in the overlap region using the asymptotic expansion of the hypergeometric function as with
$H$. In the overlap region, the form of $G$ is nearly identical to (\ref{near-H-extn}) - the only difference is in the numerical values of the constants $\tilde{c_1}$ and $\tilde{c_2}$. As before coefficient of the growing piece is never zero.
The solution for $G$ in the static limit is given by plugging $\hat \Omega = 0 $ and $D_2 = 0$ in (\ref{etazero}), since in the static limit, the equations for $G$ and $\tilde{\eta}$ decouple. The solution for $\tilde{\eta}$ is identical to that of $G$ in this limit. What is relevant is that in the asymptotic region, we are still left with an expansion for either of these functions, of the form (\ref{near-H-extn}) with $\hat \Omega = 0$, and the coefficient of the growing piece non-zero.

Let us now consider the equations for $G$ and $\tilde{\eta}$ in the general case when they are both coupled.
We split the near region into the \emph{very near horizon region} and
the \emph{far limit of the near region.}
In the \emph{very near horizon region} we use the approximation $R-1 << 1$ so that $R \sim 1$ and keep only the dominant terms with
highest powers of $(R-1)$ in the denominator in (\ref{scalarG}) and (\ref{scalarEta}). This is the region $1 < R << 2$.
The \emph{far limit of the near region} is defined by the regime where $R >> 1$, so that $(R-1)\approx R$, which can be applied to
the coupled equations (\ref{scalarG}) and (\ref{scalarEta}).
We will evaluate the solutions to the coupled equations in both regimes and present a matching argument between these two regimes in the `overlap' region $1 < R < 2$. We note that the definition of the overlap region here is not as precise as the overlap region between the near and far region in the large $n$ limit.

In the \emph{very near horizon region} approximation, the equations (\ref{scalarG}) and (\ref{scalarEta}) reduce to
\begin{subequations}\label{Sca-v-near}
\begin{align}
&\frac{d^2G}{dR^2}+\frac{dG}{dR}+\left[-\frac{3}{4}-\hat{\Omega}^2b^2\right]\frac{1}{(R-1)^2}G=\frac{4\hat{\Omega}}{(R-1)^2}\tilde{\eta}\\
&\frac{d^2\tilde{\eta}}{dR^2}+\frac{d\tilde{\eta}}{dR}+\left[-\frac{3}{4}-\hat{\Omega}^2b^2\right]\frac{1}{(R-1)^2}\tilde{\eta}=\frac{\hat{\Omega}b^2}{(R-1)^2}G.
\end{align}
\end{subequations}
These equations are still coupled. To simplify them further, we rewrite (\ref{Sca-v-near}) in terms of the
new coordinate $y=\ln (R-1)$. Also we write $G$ and $\tilde{\eta}$ as
\begin{align*}
\tilde{\eta}=e^{y/2}P \qquad G=e^{y/2}Q
\end{align*}
In the very near horizon limit $y\ra - \infty$, we now obtain
coupled differential equations with constant coefficients which can be solved analytically.
\begin{align*}
\frac{d^2Q}{dy^2}&=\lb 1+\hat{\Omega}^2b^2 \rb Q + 4\hat{\Omega}P\\
\frac{d^2P}{dy^2}&=\lb 1+\hat{\Omega}^2b^2 \rb P + \hat{\Omega}b^2Q
\end{align*}

In the case $\hat \Omega b \neq 1$, the general solutions for $\tilde{\eta}$ and $G$ are:
\begin{align}
G=&C_1(4\hat{\Omega})(R-1)^{\frac{3}{2}+\hat{\Omega}b}+C_2(4\hat{\Omega})(R-1)^{-\frac{1}{2}-\hat{\Omega}b}\nn
  &+C_3(4\hat{\Omega})(R-1)^{-\frac{1}{2}+\hat{\Omega}b}+C_4(4\hat{\Omega})(R-1)^{\frac{3}{2}-\hat{\Omega}b}\\[0.3 cm]
\tilde{\eta}=&C_1(2\hat{\Omega}b)(R-1)^{\frac{3}{2}+\hat{\Omega}b}+ C_2(2\hat{\Omega}b)(R-1)^{-\frac{1}{2}-\hat{\Omega}b}\nn
  &+C_3(-2\hat{\Omega}b)(R-1)^{-\frac{1}{2}+\hat{\Omega}b}+C_4(-2\hat{\Omega}b)(R-1)^{\frac{3}{2}-\hat{\Omega}b}
\label{Geta}
\end{align}

As expected, four arbitrary constants characterize the general solutions for these two coupled
ordinary differential equations. Let us now discuss the boundary conditions. A natural choice is finiteness of $G$ and $\tilde{\eta}$ at the horizon. However, we recall that it is the original perturbations variables $W, Y, Z$ that need to be finite at the horizon for consistency of linearized perturbation theory. If we again refer back to (\ref{Sca-def-sifi}) for the definitions of the variables $\phi, \psi, \tilde{\eta}$
in terms of $W, Y, Z$, we see that finiteness of $W,Y,Z$ at the horizon makes the choice of boundary conditions very simple. $W,Y$ are related to $G$ by a factor $(R-1)^{-\frac{1}{2}}$. This implies the following:
Regardless of the value of $\hat \Omega b$, we must set $C_2 = 0$. For $0 < \hat \Omega b < 1$, $C_3 = 0$ and $C_{1}, C_{4} \neq 0$.
For $\hat \Omega b > 1$, $C_4 = 0$ and $C_{1}, C_{3} \neq 0$.

A special case is $\hat \Omega b = 1$. The solutions for $G$ and $\tilde{\eta}$ for which the original
perturbation variables are finite at the horizon are:
\begin{eqnarray}
G = C_1 (R-1)^{\frac{5}{2}} + C_4 (4\hat{\Omega}) \sqrt{R-1}; \nonumber \\
\tilde{\eta} = C_1 \frac{b}{2} (R-1)^{\frac{5}{2}} -2 C_4 \sqrt{R-1}.
\label{specialcase}
\end{eqnarray}

Let us now look at (\ref{scalarG}),(\ref{scalarEta}) in the far limit of the near region where  we consider $R$ to be large.
In this limit $(R-1)\approx R$.
\begin{align}
\label{Sca-int-G}\frac{d^2G}{dR^2}+\frac{1}{R}\frac{dG}{dR}-\left[\frac{1}{4}+\hat{k}^2+\hat{\lambda}^2b^2+\hat{\Omega}^2b^2\right] \frac{1}{R^2}G=\frac{4\hat{\Omega}}{R^3}\tilde{\eta}\\
\label{Sca-int-eta}\frac{d^2\tilde{\eta}}{dR^2}+\frac{1}{R}\frac{d\tilde{\eta}}{dR}-\left[\frac{1}{4}+\hat{k}^2+\hat{\lambda}^2b^2+\hat{\Omega}^2b^2\right] \frac{1}{R^2}\tilde{\eta}=\frac{\hat{\Omega}b^2}{R^3}G
\end{align}

To analyze this equation, we first consider $G$ and $\tilde{\eta}$ to have a similar $R$ dependence in the large $R$ limit.
In this case we can neglect the right hand side of both the equations as it will be subleading for large $R$.
The resulting equations are decoupled and are Euler differential equations with solution
\begin{align}\label{Sca-int-sol}
G=\tilde{\eta}=a_1R^{-\frac{\sqrt{1+4\hat{\Omega}^2b^2+4(\hat{k}^2+\hat{\lambda}^2b^2)}}{2}}+
a_2 R^{\frac{\sqrt{1+4\hat{\Omega}^2b^2+4(\hat{k}^2+\hat{\lambda}^2b^2)}}{2}};
\end{align}
and similarly for $\tilde{\eta}$.

An alternative scenario in the far limit of the near region is one where  $G$ has a higher power of $R$ than $\tilde{\eta}$ in the far region. We can then neglect the right hand side in (\ref{Sca-int-G}).
The solution for $G$ will be same as (\ref{Sca-int-sol}).
We can solve for $\tilde{\eta}$ using this solution as a source term in (\ref{Sca-int-eta}). By doing this using standard
Green's function methods, we obtain a solution for $\tilde{\eta}$ which does not tally with the form of the solution for
$\tilde{\eta}$ from the far region. This indicates we cannot have this alternative scenario. The possibility of
$\tilde{\eta}$ having a higher power of $R$ than $G$ can be ruled out in a similar way.

\subsection{Matching of Solutions}

We are investigating if the black string is unstable under scalar perturbations ($\lambda, \Omega > 0$).
In the overlap region, the solution from the near region for $H$ is a growing solution with both exponents of $R$,
$R^{\frac{\sqrt{1+4\hat{\Omega}^2b^2+4(\hat{k}^2+\hat{\lambda}^2b^2)}}{2}}$ and
$R^{- \frac{\sqrt{1+4\hat{\Omega}^2b^2+4(\hat{k}^2+\hat{\lambda}^2b^2)}}{2}}$ present and the coefficient of the  growing piece always non-zero.

However, the normalizable solution from the far region is of the following form in the overlap region:
\begin{equation}
\phi=D_0 R^{-\frac{\sqrt{1+4\hat{\Omega}^2b^2+4(\hat{k}^2+\hat{\lambda}^2b^2)}}{2}};
\end{equation}
with the same exponent for $\psi$ and $\eta$ as well. Since $G$, $H$ and $\tilde{\eta}$ are sums, differences or scalar multiples
of $\phi, \psi, \eta$, the form of $H$ from the far region has only the decaying piece $R^{-\frac{\sqrt{1+4\hat{\Omega}^2 b^2+4(\hat{k}^2+\hat{\lambda}^2b^2)}}{2}}$. Thus there are no unstable modes of the type $H$ for the black
string/brane for scalar perturbations with angular momentum $l \neq 0$.

Further, in the static limit $\hat \Omega = 0$, the same matching argument can be used to conclude from (\ref{etazero}) with $D_2 = 0$ that there are no unstable modes of any type ($H$, $G$ or $\tilde{\eta}$). The same statement can also be made in the non-static case when either one of $G$ or $\tilde{\eta}$ is zero.

Let us now consider the general case when both $G$ and $\tilde{\eta}$ are non-zero. In the far limit of the near region,
$G$ is of the form (\ref{Sca-int-sol}), and a similar expression holds for $\tilde{\eta}$ since the equations decouple.
In the overlap region  $1 < R < 2$, the far region (decoupled) solution is of the form (\ref{etazero}) with arbitrary constants $D_1$ and $D_2$.
The solution for $G$ from the far limit of the near region, in the overlap region is
\begin{align}
G = &D_1(R-1)^{\frac{1}{2} + \sqrt{1 + \hat \Omega^2 b^2}}  F( \tilde p, \tilde q, 1 + 2 \sqrt{1 + \hat \Omega^2 b^2}; 1-R)+ \nn
 &D_2 (R-1)^{\frac{1}{2} -  \sqrt{1 + \hat \Omega^2 b^2}} F(1-\tilde p, 1-\tilde q, 1 - 2 \sqrt{1 + \hat \Omega^2 b^2}; 1-R).
\end{align}
The hypergeometric functions in this expression for $G$ in the overlap region are expanded as a series in $R - 1$.
Now, from the very near horizon region, the solutions for $G$ and $\tilde{\eta}$ are given by (\ref{Geta}), with finiteness of the perturbation required at the horizon.
For example, let us consider $\hat \Omega b > 1$. In the very near horizon region,
\begin{equation}
G= C_1(4\hat{\Omega})(R-1)^{\frac{3}{2}+\hat{\Omega}b}+ C_3(4\hat{\Omega})(R-1)^{-\frac{1}{2}+\hat{\Omega}b}.
\label{matchG}
\end{equation}
\begin{equation}
\tilde{\eta}=C_1(2\hat{\Omega}b)(R-1)^{\frac{3}{2}+\hat{\Omega}b}+C_3(-2\hat{\Omega}b)(R-1)^{-\frac{1}{2}+\hat{\Omega}b}.
\label{matcheta}
\end{equation}
If we consider $C_1, C_3 > 0$, then in the very near horizon region, both $G$ and $G'$ are positive. Similarly,
if both $C_1, C_3 < 0$ then both $G$ and $G'$ are negative. If $C_1> 0, C_3 < 0$, then both $\tilde{\eta}$ and
$\tilde{\eta}'$ are positive. If $C_1 < 0, C_3 > 0$, then both $\tilde{\eta}$ and
$\tilde{\eta}'$ are negative.  Thus, irrespective of the sign of the constants, either one of the two functions, $G$ or
$\tilde{\eta}$ will be such that the function and its derivative are of the same sign in the very near horizon region. For example, let this function be $G$. It needs to match with the expression for $G$ from the far limit of the
near region (\ref{etazero}) in the overlap region for some $D_1 $ and
$D_2$. We cannot match exact powers of $(R-1)$ from both sides, as this overlap region is not very precisely defined. The exponents coming from the very near horizon region depend crucially on the coupling terms, whereas in the far limit of the near region, the solutions are decoupled. However, if $\hat \Omega$ is large and $b>1$, the coupling terms will not be
significant in the overlap region. This can be seen from the equation for $G$, (\ref{scalarG}) for example, where the
coupling term is $\frac{4\hat{\Omega}}{R(R-1)^2}\tilde{\eta}$. A comparable term on the left is $\frac{\hat{\Omega}^2 b^2}{(R-1)^2}G $ which in the overlap region could be larger than the coupling term for sufficiently large $\hat{\Omega}$.
We can match features of the solutions from both sides. Let $G$ and $G'$ have the same sign from the very near horizon region. Then at leading order in $(R-1)$, in order to match this feature with (\ref{etazero}), we need $D_1 \neq 0$, since a solution with $D_1 = 0$ will have a sign opposite to its derivative.
If $D_1 \neq 0$, then in the far limit of the near region, we will have an expansion containing a piece
$R^{\frac{\sqrt{1+4\hat{\Omega}^2 b^2+4(\hat{k}^2+\hat{\lambda}^2b^2)}}{2}}$ which will not match with the solution from
the asymptotic region. Thus $G$ must be the trivial solution. Plugging this in the equation for $\tilde{\eta}$, we can conclude the same for it.

This heuristic argument is not as water-tight as the case of the perturbation $H$ due to the overlap region between the very near horizon region and the far limit of the near region not being very precisely defined. However, since the
problematic coupling terms are not significant for $\hat \Omega $ large, we do not believe there are any unstable modes of the form $G$ or $\tilde{\eta}$ at least in that case.

We can discuss the case of $\lambda$ and $\Omega$ being lower order in $n$ by setting them to zero. This is still no match
between the near and far region solutions. We cannot
set $k=0$ since we are considering non-spherically symmetric modes.

If we wish to study stable oscillatory modes with real $\omega = - i \Omega$, then in the far region, modified Bessel functions
are replaced by Bessel functions. In this case, normalizability of the perturbation allows for both the linearly independent
Bessel functions. Thus, in the overlap region, the far region solution is an arbitrary linear combination of both
$R^{\frac{\sqrt{1+4\hat{\Omega}^2b^2+4(\hat{k}^2+\hat{\lambda}^2b^2)}}{2}}$ and
$R^{- \frac{\sqrt{1+4\hat{\Omega}^2b^2+4(\hat{k}^2+\hat{\lambda}^2b^2)}}{2}}$. This can be matched to any solution from the
near region, such as one obeying quasinormal mode boundary conditions. The technique of matched asymptotic expansions yields
an approximate solution.

\section{Summary}

In this paper, we report on substantial progress in the analysis of non-spherically symmetric perturbations of the black string/flat black brane. The perturbations are decomposed in terms of the scalar, vector and tensor spherical harmonics on the $n$-sphere part of the brane metric.
By an appropriate choice of gauge, and by generalizing perturbation variables introduced by Ishibashi and Kodama for black hole perturbation theory, we have rewritten the brane perturbation equations in a vastly simplified form. The tensor perturbations which reduce to an ODE for a single function have already been discussed in \cite{kodamanotes} and do not lead to instabilities. It
is the vector and scalar perturbations that have eluded analysis before. In our formulation, the vector perturbations reduce to a system of two coupled ODEs. The scalar perturbations reduce to three coupled ODEs. To analyze stability of the black string/brane, we have assumed a time behaviour $e^{\Omega t}$ for the perturbations, and investigated if there are normalizable solutions to the perturbation equations
with $\Omega$ real and positive.

A breakthrough in the analysis of the vector and scalar perturbations comes from the use
of the large $n$ limit of general relativity \cite{kol}, \cite{emparan}. The vector equations decouple in the near-horizon
region and the asymptotic region. Due to the large $n$ limit, these regions are well-defined, and we employ the technique of matched asymptotic expansions to rule out instabilities. We require finiteness of the perturbations at the horizon for consistency of perturbation theory and normalizability asymptotically. One minor detail in the vector perturbations is that defining $\Omega = n \hat \Omega$, for $0< \hat \Omega b < 1$ ($r=b$ being the horizon location), both linearly independent
solutions to each equation are not finite at the horizon. We therefore need $\hat \Omega b \geq 1$. We do not understand if this is of any significance in the large $n$ limit. Static perturbations with $\hat \Omega = 0$ do not lead to instabilities.

Of the three scalar perturbation equations, one decouples in the near-horizon and asymptotic regions. As in the vector case, we show this does not lead to instabilities. The other two perturbations remain coupled in the near-horizon region, although they can be solved asymptotically. If any one of them is zero, the other does not lead to instabilities. In the case when both are non-zero, we employ a two step matching procedure. We split the near-horizon region into two regions with an overlap, solve the two coupled equations in the two regions, and match their features in the overlap region. We then argue that this solution does not match with the asymptotic solution, and that these perturbations cannot also lead to instability. The split of the near-horizon region into two, and the overlap region of the two is not as neat as the large $n$ split of the spacetime into a near-horizon and far region. However, for reasons outlined at the end of section IV, we believe these perturbations do not lead to instability. In the static limit $\Omega = 0$, we can show that none of the three scalar perturbations leads to an instability.

Taken together, these results in the large $n$ limit provide direct evidence from the analysis of the equations themselves that the Gregory-Laflamme instability is the only instability of the flat black brane. We have also shown that the corresponding Gross-Perry-Yaffe mode for semiclassical black hole perturbations is the \emph{unique} unstable mode in the
large $n$ limit. The above analysis can also be used to study stable perturbations with time dependence
$e^{i \omega t}$, such as quasinormal modes. In this case, the matching procedure we have outlined can be used to obtain
the quasinormal mode frequencies and approximate analytical solutions for the perturbations.

\section{Discussion}
The results of this paper have demonstrated the power of the large $D$ limit of general relativity in tackling difficult
problems in black brane perturbation theory. This has many immediate applications. As a natural next step, we are working on  the non-spherically symmetric quasinormal modes of the black string in this limit. Computing the leading order quasi-normal modes is very simple given the results of our paper. However, we need an understanding of the effect of $1/D$ corrections on the modes computed by matching the near region and far region solutions at leading order. We have not considered these corrections in our paper because we have done a stability analysis for which there are no unstable modes at the leading order and no match between the near region and asymptotic solutions.

Other projects we hope to address immediately are a study of brane perturbations with a nonzero cosmological constant, as well as an analysis of the perturbations of charged black $p$-branes considered by Gregory and Laflamme in \cite{GL1}. It was found in \cite{GL2} that the GL instability disappears for the extremal black $p$-brane. It would be interesting to know how the perturbation analysis of non-spherically symmetric perturbations of extremal black $p$-branes differs from the non-extremal ones.

Our simplified perturbation equations are amenable to a numerical study to investigate several aspects of stable black brane perturbations. We hope this work also serves as a pointer for tackling more difficult problems such as a direct study of non-spherically symmetric perturbations of curved black branes.

\appendix
 \setcounter{section}{0}
 \setcounter{subsection}{0}
 \setcounter{equation}{0}
 \renewcommand\theequation{\Alph{section}\raise.5ex\hbox{.}\arabic{equation}}

\section{Appendix A: Tensor perturbations}
For $n \geq 3$, we consider metric perturbations proportional to the tensor spherical harmonics $T_{ij}$ on $S^{n}$, which are symmetric tensors defined by $$( \hat \Delta_{n} + k_{T}^{2} ) T_{ij} = 0; ~T_{i}^{~i} = 0; ~\hat D_{j} T_{i}^{~j} = 0.$$

\begin{equation}
k_{T}^{2} = l(l + n - 1) - 2;~~~~ l= 1,2,.....
\label{a1}
\end{equation}
These metric perturbations satisfy $h_{ab} = h_{ai} = 0$; $h_{ij} = 2 r^{(4-n)/2} \Phi T_{ij}$.
For these perturbations of the black string, we assume the ansatz
$\Phi (t,r, z) = \tilde \Phi (r) e^{i \omega t} e^{i \lambda_{T} z}$. Unstable (normalizable) modes of the string correspond to imaginary values of $\omega$. The equation for $\tilde \Phi$ follows from the Einstein equations. This was derived in \cite{vstensor} for static perturbations with $\omega = 0$, and the time dependence adds only one extra term. Defining the coordinate $r_{*}$ by $dr_{*} = dr/f$, the equation for $\tilde \Phi$ is of the form
\begin{equation}
[ -\frac{d^2}{dr_{*}^{2}} + V ]\tilde \Phi = \omega^2 \tilde \Phi;
\label{a2}
\end{equation}
where $$V(r) = \frac{f}{r^2 } \left [ k_{T}^{2} + 2 n + \frac{(n^2 - 10 n + 8)}{4} + \frac{n^2}{4} \left (\frac{b}{r}\right )^{n-1} \right ] + \lambda_{T}^{2} f.$$
As can be seen, this potential is positive, and thus there are no normalizable solutions to (\ref{a2}) with $\omega$ pure imaginary. The black string is stable under this class of perturbations, and with $\omega = 0$, this also proves the semi-classical stability of the Schwarzschild-Tangherlini black holes under tensor perturbations without the need to resort to a large $n$ limit. This analysis has been done by H Kodama \cite{kodamanotes} and in the context of stability of
the Schwarzschild-Tangherlini metric under Ricci flow, in \cite{vstensor}.

Kudoh \cite{kudoh} has analyzed the tensor and vector perturbations of the black string, and a numerical analysis of scalar perturbations in an approximation. This has been done using the IK variables. However, the work suffers from serious
errors. For example, a cross-check is that the eigenvalue equation (\ref{a2}) must reduce to the linearized Einstein equations on Schwarzschild-Tangherlini black holes for $\lambda=0$, given in Ishibashi and Kodama's paper \cite{rev} as well as older work of Gibbons and Hartnoll \cite{gibhart}. It indeed does. However, Kudoh's equation for tensor perturbations, given by equation(39) in \cite{kudoh}, does not match (\ref{a2}) and does not reduce to the relevant equations of
Ishibashi and Kodama, or Gibbons and Hartnoll upon setting $\lambda=0$. We believe the likely errors are typos in the coefficients of the $f$ and $f'$ terms in equation (39) in \cite{kudoh}. Similarly for the vector case, it is claimed in \cite{kudoh} that the equations for $(F_r,F_t)$ completely decouple. We explicitly show that the same equations do not decouple even in the large $n$ limit. Our equations in the vector case match those of H Kodama in \cite{kodamanotes} --- Kodama uses
gauge invariant variables on the brane, and we use a gauge-fixed formalism. However, this only results in an extra constraint and equation in Kodama's case. Thus we believe that there are errors in the analysis of Kudoh \cite{kudoh} that invalidate many of the claims in that paper.
\section{Appendix B: Handling source terms}
Here, we take as an example, the first of the equations (\ref{Brhs}), which is given by
\begin{equation}
\frac{d^2A}{dr^2}+\frac{n}{r}\frac{dA}{dr}+\left(-\frac{k_v^2}{r^2}-\lambda^2-\Omega^2\right)A=\left(\frac{2}{r}\right)\Omega B .
\label{sourcea1}
\end{equation}
In sub-section (III.2), $B$ was evaluated and found to be $r^{-(n-1)/2} K_\nu (\nu z)$ where $z = \frac{ \sqrt{\lambda^2+\Omega^2}}{\nu} r$. Let us take $A = r^{-(n-1)/2} S$. Then the equation (\ref{sourcea1}) becomes a modified Bessel equation with source terms. This is
\begin{equation}
\frac{d^2 S}{dz^2} + \frac{1}{z} \frac{dS}{dz} - [1 + \frac{\nu^{2}}{z^2} ] S =
 \frac{2 \hat \Omega}{\sqrt{\hat \lambda^2 + \hat \Omega^2} z} K_\nu (\nu z).
\label{sourcea2}
\end{equation}
The Wronskian of the two linearly independent solutions to the homogeneous equation
$W[I_\nu (\nu z), K_\nu (\nu z)] = \frac{1}{z}$. We use the method of variation of parameters to write the solution to
(\ref{sourcea2}). This takes the form (replacing $dz = \frac{ \sqrt{\lambda^2+\Omega^2}}{\nu} dr$)
\begin{equation}
S = - \frac{2 \hat \Omega}{\nu} I_\nu (\nu z) \int (K_\nu (\nu z) )^2  ~dr  ~+~
\frac{2 \hat \Omega}{\nu} K_\nu (\nu z) \int K_\nu (\nu z) I_\nu (\nu z)   ~dr.
\label{sourcea3}
\end{equation}

Inserting the asymptotic expansions for modified Bessel equations for large argument and order, we note that
as $r \to \infty$, $S \to 0$ exponentially as $e^{-\nu z}$. In the overlap region, we can change variables from $r$ to $R$ using (\ref{rR}) and
$K_\nu (\nu z) I_\nu (\nu z) \sim [1 + (\frac{\lambda^2+\Omega^2}{\nu^2}) b^2 ]^{-1/2}$. We have $dr \sim dR/R$, and from Appendix D we observe that in the large $n$ approximation in the overlap region, $K_\nu (\nu z) = c R^{-d}$ where $c$ is a constant and $d = \frac{\sqrt{1+4\hat{\Omega}^2 b^2+ 4(\hat{k}_v^2+\hat{\lambda}^2b^2 )}}{2}$. By a similar computation,
$I_\nu (\nu z) = \tilde c R^{d}$. Using this, we evaluate (\ref{sourcea3}) in the overlap region to obtain
$S = (const.) R^{-d}$, and  $A = r^{-(n-1)/2} S = (const.) R^{-1/2 -d}$.

\section{Appendix C: Equations for Scalar Perturbations}

\renewcommand\theequation{\Roman{equation}}
In this section we give an outline of the procedure to get equations in terms of $W,Y,Z$ variables. We first write equations $2\delta G_{\mu\nu}=-\lambda^2 h_{\mu\nu}$ in terms of these.

Equation for $\delta G_{ti}$ :
\begin{equation}\label{f_t}
\partial_t W + \partial_r Z = \lambda^2 r^{n-2} \left[X_t +\frac{1}{k^2}\partial_t(r^2H_T)\right]
\end{equation}

Equation for $\delta G_{ri}$ :
\begin{equation}\label{f_r}
\partial_r Y+\frac{f'}{2f}Y-\frac{f'}{2f}W -\frac{1}{f^2}\partial^2_t Z = \lambda^2r^{n-2}\left[X_r+\frac{1}{k^2}\p_r(r^2H_T)\right]
\end{equation}

Equation for $\delta G_{t}^r$ :
\begin{align}\label{f_rt}
&\left[\frac{k^2}{r^2}-f'' -\frac{nf'}{r} \right]Z
    + f\partial_t\partial_r Y
   + \left(\frac{2f}{r}-\frac{f'}{2}\right) \partial_t Y  \nonumber \\
 &+ f\partial_t\partial_r W
  - \left(\frac{(n-2)f}{r}+\frac{f'}{2}\right) \partial_t W =-\lambda^2 r^{n-2}f^r_t \nonumber \\& - \frac{2\lambda^2}{nk^2}\left[n\p_t\p_r(r^2H_T)-\frac{nf'}{2}\p_r(r^2H_T)\right]
\end{align}

Equation for $\delta G_{r}^r$ :
\begin{align}\label{f_rr}
& \frac{1}{f}\partial_t^2 W -\frac{f'}{2}\partial_r W +\frac{1}{f}\partial_t^2 Y -\left(\frac{f'}{2}+\frac{nf}{r}\right)
   \partial_r Y \notag\\
& +\left[\frac{n-1}{r^2}(f-1)+\frac{(n+2)f'}{2r}+\frac{f''}{n}\right]W \notag\\
& +\left[\frac{1-f}{r^2}-\frac{3n-2}{2r}f'-\frac{n-1}{n}f''+\frac{k^2-nK}{r^2}\right]Y
    \notag\\
& + \frac{2n}{rf}\partial_t Z = -\lambda^2 r^{n-2}f^r_r \notag \\
& -\frac{2\lambda^2}{nk^2}\left[-\frac{nk^2}{r^2}(H_Tr^2)-\frac{n}{f}\p^2_t(r^2H_T)+\frac{nf'}{2}\p_r(r^2H_T)+\frac{n^2f}{r}\p_r(r^2H_T)\right]
\end{align}

Equation for $\delta G_i^i$ :
\begin{align}\label{f_L}
& \frac{1}{2f} \partial_t^2 W +\frac{f'}{4} \partial_r W -\frac{f}{2} \partial_r^2 Y
   -\left(\frac{3f'}{4}+\frac{f}{r} \right)\partial_r Y \notag\\
& +\left[ \frac{(n-1)(n-2)(f-1)}{2nr^2}
   +\frac{(6n-4-n^2)f'}{4nr} +\frac{f''}{2n} \right] W \notag\\
& +\left[ \frac{(n-1)(n-2)(f-1)}{2nr^2}
    +\frac{(-n^2+2n-4)f'}{4nr}-\frac{(n-1)f''}{2n}\right] Y \notag\\
& +\left(\frac{1}{rf}-\frac{f'}{2f^2}\right) \partial_t Z
   +\frac{1}{f}\partial_t\partial_r Z=-\lambda^2 r^{n-2}H_L -\frac{\lambda^2}{nk^2}\bigg[\frac{(n-1)k^2}{r^2}(H_Tr^2)\nn &-nf\p^2_r(r^2H_T)+\frac{n}{f}\p^2_t(r^2H_T)-nf'\p_r(r^2H_T)-\frac{n(n-1)f}{r}\p_r(r^2H_T)\bigg]
\end{align}

Equation for $\delta G_{t}^t$ :
\begin{align}\label{f_tt}
 & -f\partial_r^2 W+ \left(\frac{n-4}{r}f-\frac{f'}{2}\right)
    \partial_r W -f\partial_r^2 Y - \left(\frac{f'}{2}+\frac{4f}{r}\right)
    \partial_r Y  \notag\\
& -\left[\frac{n-1}{r^2}-\frac{(2n-3)f}{r^2}
    +\frac{n-2}{2r}f'+\frac{n-1}{n}f''
    -\frac{k^2}{r^2}\right]W \notag\\
& -\left[\frac{n-1}{r^2}-\frac{n-3}{r^2}f
   + \frac{(n-2)f'}{2r}
   -\frac{f''}{n}\right] Y =-\lambda^2 r^{n-2}f^t_t \notag \\
&  -\frac{2\lambda^2}{nk^2}\left[-\frac{nk^2}{r^2}(H_Tr^2)+nf\p^2_r(r^2H_T)+\frac{nf'}{2}\p_r(r^2H_T)+\frac{n^2f}{r}\p_r(r^2H_T)\right]
\end{align}

As discussed before, the right-hand side of these equations have components of $(f_{ab},X_a,H_L,H_T)$. To get them in terms of $W,Y,Z$, we have to combine these equations. Let us expand our variables in terms of these components.

\begin{align*}
& \frac{W}{r^{n-2}} = f^t_t-\frac{2}{f}\p_tX_t+\left(f'-\frac{2f}{r}\right)X_r-2H_L-\frac{2H_T}{n}\\
& \frac{Y}{r^{n-2}} = f^r_r+2f\p_rX_r+\left(f'-\frac{2f}{r}\right)X_r-2H_L-\frac{2H_T}{n}	\\
& \frac{Z}{r^{n-2}} = f^r_t+f\p_rX_t+f\p_tX_r-f'X_t
\end{align*}

\setcounter{equation}{0}
\renewcommand\theequation{\alph{equation}}
The final $W,Y$ and $Z$ equations are as follows:
Looking at the expression of $Z$, we see that adding (\ref{f_rt}), and derivatives of (\ref{f_t}) and (\ref{f_r}) with appropriate coefficients will give the right-hand side of the resulting equation in terms of the $Z$ variable.
\begin{align}\label{Z}
&\partial_r^2 Z -\frac{1}{f^2}\partial_t^2 Z-\left(\frac{(n-2)}{r}+\frac{f'}{f}\right)\partial_r Z - \nonumber\\ &\left[\frac{k^2}{f r^2}-\frac{f''}{f}-\frac{nf'}{f r}\right]Z-\left(\frac{2}{r}-\frac{f'}{f}\right)\partial_t Y-\frac{f'}{f}\partial_t W = \frac{\lambda^2}{f}Z
\end{align}

Similarly adding (\ref{f_rr}), the derivative of (\ref{f_r}) and (\ref{f_L}) will give us the equation for $Y$.
\begin{align}
\label{Y}&\partial_r^2 Y-\frac{1}{f^2}\partial_t^2 Y-\left(\frac{n}{r}-\frac{f'}{f}\right)\partial_r Y- \nonumber \\
&\left[-\frac{2(n-1)}{nr^2f}+\frac{-n^2+2n-2}{nr^2}+\frac{(2-n)}{nrf}f'-\frac{f''}{f}+\frac{f'^2}{2f^2}+\frac{k^2}{r^2f}\right]Y -\nonumber\\
&\left[-\frac{2(n-1)}{nr^2f}+\frac{2n-2}{nr^2}+\frac{2-n}{nrf}f'+\frac{f''}{f}-\frac{f'^2}{2f^2}\right]W+\frac{2f'}{f^3}\partial_t Z = \frac{\lambda^2}{f}Y
\end{align}

To obtain the equation for $W$, we add (\ref{f_tt}), the derivative of (\ref{f_t}), (\ref{f_L}) and (\ref{f_r}).
\begin{align}
\label{W}&\partial_r^2 W-\frac{1}{f^2}\partial_t^2 W-\left(\frac{(n-4)}{r}-\frac{f'}{f}\right)\partial_r  W- \nonumber \\ &\left[-\frac{2(n-1)}{nr^2f}+\frac{n^2-2}{nr^2}+\frac{(2-3n)}{nrf}f'-\frac{f''}{f}+\frac{f'^2}{2f^2}+\frac{k^2}{r^2f}\right]W-\nonumber\\
&\left[-\frac{2(n-1)}{nr^2f}-\frac{2}{nr^2}+\frac{2+n}{nrf}f'+\frac{f''}{f}-\frac{f'^2}{2f^2}\right]Y-\left(\frac{2f'}{f^3}-\frac{4}{rf^2}\right)\partial_t Z = \frac{\lambda^2}{f}W
\end{align}

We see that the extra $H_T$ terms automatically vanish and the equations are coupled. In the static limit, the $Z$ equation decouples and we get coupled equations for $(W,Y)$.

\setcounter{equation}{0}
\section{Appendix D: Expansion of Modified Bessel Functions}

In this section, we write the expansion of $K_\nu(\sqrt{\lambda^2+\Omega^2} r)$ in terms of $R$.
In the modified Bessel function $K_\nu(\sqrt{\lambda^2+\Omega^2} r)$ , both the Bessel function order $\nu$ and argument $\sqrt{\lambda^2+\Omega^2} r$ are of same order in $n$. As we are working in the large $n$ approximation, we have to use
expansions for modified Bessel functions of large order and large argument.

Let us denote $\kappa=\sqrt{\lambda^2+\Omega^2}$. For simplicity of calculation, we define a new coordinate $z=\kappa r/\nu $. Here
\begin{equation}
\nu=\sqrt{\frac{n^2+1}{4}+k^2}\approx\frac{n}{2}\sqrt{1+4\hat{k}^2}
\end{equation}

The far region solution (\ref{far-phi-sol}) can be now written as
\begin{equation*}
\phi=D_2\frac{\nu}{\kappa}\sqrt{z} K_\nu(\nu z)
\end{equation*}

The large order and large argument expansion of this expression is,
\begin{equation}
\sqrt{z} K_\nu(\nu z)=
\sqrt{\frac{\pi}{2\nu}}\frac{\sqrt{z}}{(1+z^2)^{1/4}}e^{-\nu\eta}\left[1+\sum_{m=1}^{\infty}(-1)^m\frac{u_m(\tilde{t})}{\nu^m}\right]
\end{equation}

where
\begin{align*}
&\eta=\sqrt{1+z^2}+\ln\left[\frac{z}{1+\sqrt{1+z^2}}\right]
&\tilde t=\frac{1}{\sqrt{1+z^2}}
\end{align*}

and $u_m(\tilde t)$ are polynomials in $\tilde t$. We are only considering terms to highest order in $n$. We ignore the polynomial terms as they are divided by $\nu$. Substituting for $\eta$ we get, up to a constant,

\begin{equation}\label{z_exp}
\sqrt{z} K_\nu(\nu z)\approx\frac{1}{z^\nu}\left(1+\sqrt{1+z^2}\right)\frac{\sqrt{z}}{(1+z^2)^{1/4}}\exp\left[-\nu\sqrt{1+z^2}\right].
\end{equation}

In order to express $K_\nu(\nu z)$ in the overlap region, we write this expression in terms of $R$. Here, $R=\frac{r^{n-1}}{b^{n-1}}$. To expand the expression in orders of $n$, we use the following definition of $r$ in terms of $R$, valid in the overlap region,
\begin{equation*}
r=b \left[1+\frac{\ln R}{n-1}\right].
\end{equation*}

We will now look at each term in (\ref{z_exp}) individually. For $z^\nu$, the term is directly proportional to $r^n$. Hence we use the definition $r^n=b^n R$ for large $n$.
\begin{equation}
\frac{1}{z^\nu}=\left(\frac{\nu}{\kappa}\right)^\nu \frac{1}{r^\nu}=\left(\frac{\nu}{\kappa}\right)^\nu b^{-\nu}R^{-\frac{\nu}{n}}=\left(\frac{\nu}{\kappa b}\right)^\nu R^{-\frac{\sqrt{1+4\hat{k}^2}}{2}}
\end{equation}

The next term becomes

\begin{align}
\left[1+\sqrt{1+z^2}\right]^\nu=&\exp\left[\nu\ln\left\lbrace 1+ \lb 1 + \frac{\kappa^2b^2}{\nu^2}\lb(1+2\frac{\ln R}{n}\rb\rb^{1/2}\right\rbrace\right]\nn
=& \lb1+\sqrt{1+\frac{\kappa^2b^2}{\nu^2}}\rb^\nu \exp\ltb \frac{\kappa^2b^2\sqrt{1+\frac{\kappa^2b^2}{\nu^2}}}{\nu n \lb 1+\sqrt{1+\frac{\kappa^2b^2}{\nu^2}}\rb} \ln R\rtb .
\end{align}

$\kappa$ and $\nu$ are of order $n$. Hence the constant multiplying $\ln R$ is of order $1$.
Similarly substituting for $z$ we get,
\begin{equation}
\exp\ltb -\nu\sqrt{1+z^2}\rtb = \lb 1+\frac{\kappa^2b^2}{\nu^2}\rb^{-\nu/2}\exp\ltb \frac{-\kappa^2b^2}{n\nu\sqrt{1+\frac{\kappa^2b^2}{\nu^2}}}\ln R\rtb
\end{equation}

The coefficient of $\ln R$ is of order $1$ in this term.
The remaining term in (\ref{z_exp}) becomes,
\begin{align}
\frac{\sqrt{z}}{(1+z^2)^{1/4}}= \lb 1+ \frac{\nu^2}{\kappa^2b^2}\rb^{-1/4}\exp\ltb\frac{\nu^2}{2n\kappa^2b^2}\lb 1+ \frac{\nu^2}{\kappa^2b^2}\rb^{-1} \ln R \rtb
\end{align}

Here, the constant multiplying $\ln R$ is of order $1/n$. Therefore, this term is sub-leading in comparison with the other terms in expansion. We are interested in terms that are leading order in $n$. In the final expression, we neglect this term. Substituting all the expressions in $R$ back in (\ref{z_exp}), we get the following expression for $\phi$ (we have absorbed all the constants in $D_0$) :

\begin{equation}
\phi=D_0 R^{-\frac{\sqrt{1+4\hat{\Omega}^2b^2+4(\hat{k}^2+\hat{\lambda}^2b^2)}}{2}}
\end{equation}

Notice that in the final expression, we have neglected terms coming from $\sqrt{z}$ as they are sub-leading. Hence we can write

\begin{equation}\label{K_large_n}
K_\nu(\nu z)= (const.)R^{-\frac{\sqrt{1+4\hat{\Omega}^2b^2+4(\hat{k}^2+\hat{\lambda}^2b^2)}}{2}}
\end{equation}

Expansion for the modified Bessel function of first kind $I_\nu(\nu z)$ for large order and large argument is
\begin{equation}
I_{\nu}(\nu z)=\frac{1}{\sqrt{2\pi\nu}}\frac{1}{(1+z^2)^{1/4}}e^{\nu\eta}\left[1+\sum_{m=1}^{\infty}\frac{u_m(\tilde{t})}{\nu^m}\right]
\end{equation}

We can obtain expansion of $I_\nu(\nu z)$ in the overlap region in the large $n$ approximation by replacing $(-\nu)$ by $\nu$ in the expansion of $K_\nu(\nu z)$. The final expression for $I_\nu$ is
\begin{equation}\label{I_large_n}
I_\nu(\nu z)= (const.)R^{\frac{\sqrt{1+4\hat{\Omega}^2b^2+4(\hat{k}^2+\hat{\lambda}^2b^2)}}{2}}
\end{equation}

\end{document}